\documentclass[pra,twocolumn,amssymb]{revtex4}
\usepackage{graphicx}
\usepackage{bm}
\usepackage{epsf}
\usepackage{amssymb,amsfonts}
\usepackage{times}
\usepackage{graphicx}
\usepackage{bm}
\usepackage{amsmath}
\usepackage{ulem}
\usepackage[usenames]{color}

\def \be {\begin{equation}}
\def \ee {\end{equation}}
\def \p {\partial}
\def \BEA {\begin{eqnarray}}
\def \EEA {\end{eqnarray}}
\def \BC {\begin{cases}}
\def \EC {\end{cases}}

\begin{document}

\title{Spin-charge separation in an Aharonov-Bohm interferometer}
\author{A. P. Dmitriev$^{1}$}
\author{I. V. Gornyi$^{1,2}$}
\author{V. Yu. Kachorovskii$^{1,2,3}$}
\author{D. G. Polyakov$^{2}$}

\affiliation{
$^{1}$A.F.~Ioffe Physico-Technical Institute, 194021 St.Petersburg, Russia\\
$^{2}$Institut f\"ur Nanotechnologie, Karlsruhe Institute of Technology, 76021 Karlsruhe, Germany\\
$^{3}$L.D.~Landau Institute for Theoretical Physics, 119334 Moscow, Russia
}

\begin{abstract}
We study  manifestations of spin-charge separation (SCS) in transport through a tunnel-coupled interacting single-channel quantum ring. We focus on the high-temperature case (temperature $T$ larger than the level spacing $\Delta$) and discuss both the classical (flux-independent) and interference  contributions to  the tunneling conductance of the ring in the presence of magnetic flux. We demonstrate that the SCS effects, which arise solely from the electron-electron interaction, lead to the appearance of a peculiar fine structure of the electron spectrum in the ring. Specifically, each level splits into a series of sublevels, with their spacing governed by the interaction strength. In the high-$T$ limit, the envelope of the series contains of the order of $T/\Delta$ sublevels. At the same time, SCS suppresses the tunneling width of the sublevels by a factor of $\Delta/T$. As a consequence, the classical transmission through the ring remains unchanged compared to the noninteracting case: the suppression of tunneling is compensated by the increase of the number of tunneling channels. On the other hand, the flux-dependent contribution to the conductance depends on the interaction-induced dephasing rate which is known to be parametrically increased by SCS in an infinite system. We show, however, that SCS is not effective for dephasing in the limit of weak tunneling. Moreover, generically, in the almost closed ring, the dephasing rate does not depend on the interaction strength and is determined by the tunneling coupling to the leads. In certain special symmetric cases, dephasing is further suppressed. Similar to the spinless case, the high-$T$ conductance shows, as a function of magnetic flux, a sequence of interaction-induced sharp negative peaks on top of the classical contribution.
\end{abstract}

\maketitle

\section{Introduction}
\label{s1}

Spin-charge separation (SCS) is a hallmark of non-Fermi-liquid behavior \cite{solyom79,voit94,gogolin98,giamarchi04}. The essence of SCS is that {\it single-electron} excitations factorize in space-time into two parts which independently exhibit dynamics of, respectively, the spin and charge degrees of freedom. In one-dimensional (1D) systems, SCS is inherently linked to the decoupling of two types of elementary {\it bosonic} spin and charge density excitations which separately carry either spin or charge and propagate with different velocities, $v$ and $u$, respectively. Experimental evidence of SCS in nanowires was demonstrated in electron tunneling \cite{auslaender05,jompol09}, thermal transport \cite{Lorenz02}, and, more recently, in spin-filtering \cite{hashisaka17} experiments. The effect of SCS is most pronounced in the ``spin-incoherent regime'' \cite{fiete07,matveev04,fiete04,cheianov04,hew08} which is realized in 1D systems with strongly different spin and charge velocities.

\begin{figure}[ht!]
\leavevmode \epsfxsize=6cm
\centering{\epsfbox{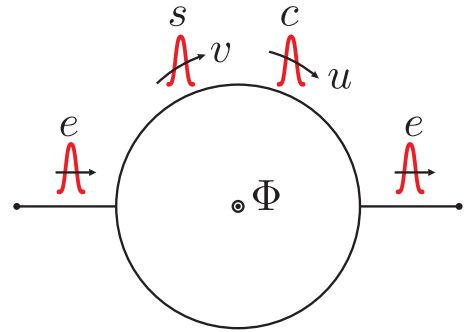}}
\caption{Quantum ring threaded with magnetic flux $\Phi$. An electron ($e$) that tunnels from the left Fermi lead to the ring splits into the charge ($c$) and spin ($s$) excitations propagating with different velocities $u$ and $v$, respectively. The factorized spin and charge parts of the electron combine again at the tunnel contact to the right Fermi lead, possibly after many rotations, in order that the electron be able to pass through the ring and escape to the right lead.}
\label{fig1}
\end{figure}

One of the key problems that are related to SCS is about the manifestation of SCS in the quantum interference of electron waves. This is the subject of the present paper. We focus on the, perhaps, conceptually simplest device for specifically probing the interference---a single-channel quantum ring tunnel-coupled to the leads (see Fig.~\ref{fig1}) and threaded with the magnetic flux $\Phi$. The conductance of the ring ${\rm G}(\phi)$ exhibits the Aharonov-Bohm (AB) effect \cite{bohm59,aronov87}, i.e., changes periodically with the dimensionless magnetic flux $\phi=\Phi/\Phi_0$, where $\Phi_0=hc/e$ is the flux quantum,---with a period $1$---because of the interference of electron trajectories winding around the hole. The sensitivity of the phase of an electron wavefunction to the flux enables the design of AB interferometers
\cite{aronov87,AB1,yacoby96,AB2,bykov00,bykov00a,AB3,AB4,AB5,AB6,AB7,AB8,AB9,roulleau07,roulleau08,zhang09,weisz12} that can be tuned by the external magnetic field. The peculiar predictions that we make in this paper appear to be amenable to experimental verification on many-electron nanorings, with a few or single conducting channels, which have already been produced \cite{shea00,piazza00,fuhrer01,keyser03,zou07}.

We consider a clean ring without disorder, so that electrons only experience scattering on the contacts---and because of interactions with each other. Consider first the noninteracting limit. Transmission of an electron through the ring can occur along paths with different numbers and different sequences of clockwise and anticlockwise windings between the left and right contacts, characterized by different transmission amplitudes $A_i$, where $i$ is the index for a particular path. The classical and interference  contributions to the transmission coefficient are then proportional to $\sum_i |A_i |^2$ and $\sum_{i\neq j} A_i A_j^*$, respectively. The classical contribution does not depend on $\phi$, so that an efficient manipulation of the AB interferometer by the external magnetic field relies on the existence of the interference contribution.

A key obstacle hindering the AB interference is interaction-induced dephasing of the electron waves. In the presence of interactions, electrons on the ring form a Luttinger liquid (LL) in the ground state, with SCS being one of the inherent properties of the spinful LL. A ``natural expectation" would be that SCS enhances, possibly strongly, dephasing of the AB oscillations. Indeed, the single-electron excitation that is created in the ring after tunneling from the lead splits into the charge and spin components which start propagating with different velocities, as illustrated in Fig.~\ref{fig1}. This decomposition of an electron into spatially separated charge and spin pieces is known to increase the decay rate of single-electron excitations in an infinite system to a value of the order of $\alpha T$ (see, e.g., Refs.~\cite{gornyi05,lehur05,yashenkin08}), compared to the decay rate of the order of $\alpha^2 T$ for the spinless case. Here, $\alpha>0$ is the dimensionless constant characterizing the strength of repulsive interaction, which we assume to be small, $\alpha \ll 1$
 (experimentally, the value of $\alpha$ is controlled by  the electrostatic environment; in particular,  by the distance to the metallic gate).  For spinful electrons, the same decay rate governs dephasing of the quantum interference conductivity correction in an infinite disordered Luttinger liquid \cite{yashenkin08}. One might thus expect a similar enhancement of dephasing by SCS in the spinful ring.

In the present paper, we show that, in fact, in the weakly tunnel-coupled interferometers, the ``AB dephasing rate'' $\Gamma_\varphi$ is insensitive to SCS.
Moreover, generically, $\Gamma_\varphi$ does not depend on the interaction strength and is determined by the tunneling coupling to the leads. In certain special symmetric cases, dephasing is further suppressed. In particular, this happens for the case of fully isotropic (in spin and chirality spaces) interaction.

To an extent, the insensitivity of $\Gamma_\varphi$ to the strength of generic interactions in the weak-tunneling limit is similar to the spinless case. Indeed, apart from Ref.~\cite{yashenkin08}, our approach to the SCS effects in the ring geometry builds upon the earlier work on transport of spinless electrons through a quantum ring \cite{dmitriev10,dmitriev14}. As was shown there, the dephasing rate in an almost closed spinless ring is given by the total tunneling rate, namely the rate at which the ring exchanges electrons with the leads. In the present work, we find that, generically, the dominant mechanism of dephasing in the spinful ring is the same---the so-called zero-mode (ZM) dephasing \cite{dmitriev10}. This, in turn, means that $\Gamma_\varphi$ is determined by the  total tunneling rate as in the spinless case, thus vanishing in the limit of weak tunneling. We also find that this mechanism is not effective in the symmetric cases mentioned above.

There is one more point of similarity---now only at the qualitative level---between the spinless and spinful systems: we show that the destructive interference between right- and left-moving electrons leads, in the presence of SCS, to a series of sharp interaction-induced negative peaks in the high-$T$ conductance as a function of magnetic flux, bearing resemblance to the interference pattern in the spinless case \cite{dmitriev10}. Importantly, however, the width and depth of the envelope of the AB conductance peaks are strongly modified by SCS.

To be more specific, it is useful to recall the origin of the interference pattern in the high-$T$ limit in the case of spinless electrons. To begin with, for {\it noninteracting} electrons (on a disorder-free ring weakly tunnel-coupled to the contacts), the sharp antiresonances in the function $G(\phi)$ in the high-$T$ limit occur at $\phi = 1/2+n$, where $n$ is integer \cite{imry}. The antiresonances originate from the destructive interference in tunneling via pairs of quantum levels inside the ring for electrons of opposite chirality. At $\phi=1/2$, the levels of electrons rotating clockwise and anticlockwise are pairwise exactly degenerate and the tunneling amplitudes for two levels in each pair are of opposite sign. The total transmission coefficient at $\phi=1/2$ is thus exactly zero for an arbitrary energy of the tunneling electron, which explains the survival of the interference pattern as $T$ increases in the noninteracting case. Electron-electron interactions change the picture dramatically. As shown in Ref.~\cite{dmitriev10}, interactions between spinless electrons of opposite chirality can be incorporated into an effective magnetic flux dependent on the circular current inside the ring. Tunneling-induced fluctuations of the circular current and, in turn, of the effective flux split the antiresonance at $\phi=1/2$ into a series of peaks (``persistent-current blockade''). This is the ``interference pattern" for spinless electrons that we referred to in the above.

The picture based on the introduction of the effective flux controlled by the circular current \cite{dmitriev10} is no longer valid in the presence of SCS. It is thus the fate of the persistent-current blockade in the presence of the spin degree of freedom that is one of the subjects of this paper. As already mentioned above, SCS does not wipe out the splitting of the ``noninteracting" antiresonance in $G(\phi)$ into a series of sharp peaks. Rather, SCS brings about new physics behind the emergence of the resonant structure and, consequently, modifies its parameters.  Moreover, SCS determines the characteristic transparency of the tunnel contacts at which the fine structure in ${\rm G}(\phi)$ blurs out as the tunneling rate is increased: this occurs when the dephasing rate $\Gamma_\varphi$ becomes of the order of the ``spin-charge collision rate" $1/\tau_{\rm sc}$, at which the paths of the spin and charge components cross each other. For larger $\Gamma_\varphi$, all resonances overlap and form a single dip in ${\rm G}(\phi)$ with a width given by the single-particle decay rate $\alpha T$ in units of the level spacing in the absence of interaction $\Delta$.

\begin{figure}
\leavevmode \epsfxsize=5cm
\centering{\epsfbox{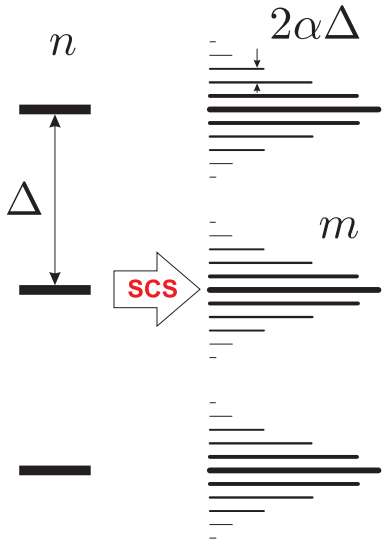}}
\caption{SCS induces splitting of each electron level of a noninteracting ring (enumerated by index $n$) into a series of sublevels (index $m$). The characteristic number of the sublevels that have a large contribution to the tunneling amplitude for given $n$ is of the order of $T/\Delta$. The sublevels have different widths  which decrease from the center of the group, as schematically illustrated by lines of different thickness. The distance between the nearest sublevels is proportional to the interaction strength.}
\label{F-new}
\end{figure}

Another nontrivial result of this work is for the classical part of the tunneling conductance. In a simpleminded approach to the problem, one would think that also the classical transmission through the tunnel-coupled ring is suppressed because of SCS---indeed, for essentially the same reason as in the case of the interference term in the transmission coefficient. The rationale would be that, the tunneling escape from the ring can only happen if the spin and charge components of the single-electron excitation collide in the vicinity of the contacts to the lead, and these collisions are rare (see Fig.~\ref{fig1}). We show that this expectation is not true, either---in fact, independently of the strength of tunneling. The subtle point is that the electron-electron interaction between {\it spinful} electrons splits each level in the ring into a series of sublevels. This is a direct consequence of SCS. As we demonstrate below, although the tunneling width of the {\it sublevels} is indeed suppressed by SCS, this effect is compensated in the conductance by the increase of the number of tunneling channels, as illustrated in Fig.~\ref{F-new}.

Let us clarify the last point in more detail. The key ingredient of the underlying physics here is the multiple windings and, consequently, multiple returns to the contacts in the finite-size system. The characteristic time between spin-charge collisions near the contact $\tau_{\rm sc}$ is given for $\alpha\ll 1$ by $\tau_{\rm sc}=2\pi v/(u-v)\Delta\simeq\pi/\alpha\Delta$ (the difference between the charge and spin velocities, $u$ and $v$, is linear in $\alpha$ for small $\alpha$). For $T\gg\Delta$, the characteristic ``dwelling time" $\tau_{\rm d}$ during which the spin and charge excitations run together as a whole (and, therefore, can tunnel out of the ring) is much shorter and given by $\tau_{\rm d}=2\pi v/(u-v)T\simeq\pi/\alpha T$ (note a similarity between $\tau_{\rm d}$ and the electron lifetime in an infinite system). As a consequence, the tunneling rate is suppressed by the factor $\tau_{\rm d}/\tau_{\rm sc}\simeq\Delta/T$, independent of $\alpha$ for $\alpha\ll 1$. On the other hand, as we show below, the spin-charge collisions are, in essence, correlated even if the spin and charge velocities are not commensurate, namely the spin and charge collide {\it periodically}, with a period given by $\tau_{\rm sc}$. It is this periodicity that leads to the formation of a fine structure in the electron spectrum and to the resulting increase of the number of tunneling channels by a factor of $T/\Delta$ (Fig.~\ref{F-new}). As a consequence, the classical transmission coefficient through the ring does not change, and nor does the dephasing rate for the AB oscillations.

Note that the emergence of the SCS-induced fine structure of AB resonances was discussed earlier in Refs.~\cite{jagla93,hallberg04,meden08,rincon09} for strongly interacting electrons, with emphasis on the effects of commensurability between $v$ and $u$. In particular, Refs.~\cite{hallberg04} and \cite{rincon09} considered the ``spin-incoherent" limit of the $t$-$J$ model (which corresponds to the limit of a strong onsite Hubbard repulsion) for a quantum ring made of a finite number of sites filled with $N$ particles, with the ratio $v/u$ being small in the parameter $1/N$ but finite. The suppression of transport through the ring for certain $\phi$ in the limit of strong Hubbard interactions was associated in Ref.~\cite{rincon09} with level crossings in the ground state of $N$ particles. By contrast, in our continuous Luttinger-liquid model, similar---in this respect---to that of Refs.~\cite{jagla93} and \cite{meden08}, the $1/N$ effects are irrelevant for the relation between $v$ and $u$---and altogether for the emergence of the SCS-induced structure in the AB resonances.

It is worth noting that Refs.~\cite{jagla93,hallberg04,meden08} presented their results in terms of the zero-$T$ conductance averaged over the period of the energy spectrum.  Although such a quantity shows interaction-induced splitting of the resonances, this approach does not produce a solution of the finite-$T$ problem, not even in the high-$T$ limit, where typical excitations have energies much larger than the characteristic level spacing. This is because it does not include important aspects of the finite-$T$ dynamics of the system, namely the interaction-induced decay of single-particle excitations at finite $T$ and the effect of dephasing. Below, we develop an analytical theory for the high-$T$ AB conductance, taking account for both effects.

The paper is organized as follows. Section~\ref{s2} covers some of the basic aspects of SCS in an infinite system (Sec.~\ref{s2a}), in the isolated ring (Sec.~\ref{subsec:IIB}), and in the ring tunnel-coupled to the leads (Sec.~\ref{s2c}). In Sec.~\ref{s3}, we discuss the two-particle dynamical properties of the spin-charge separated ring in the absence (Sec.~\ref{s3a}) and presence (Sec.~\ref{s3b}) of tunneling. In Sec.~\ref{s4}, we calculate the classical (Sec.~\ref{s4a}) and interference (Sec.~\ref{s4c}) contributions to the conductance, and the dephasing rate that governs the latter (Sec.~\ref{sec:deph}), with the main results for the case of isotropic interactions summarized in Sec.~\ref{s4c1}. Our conclusions are presented in  Sec.~\ref{s5}. Some of the technical details are placed in the Appendixes.

\section{Basics}
\label{s2}

Below, we study the linear response conductance of the AB interferometer which consists of a spinful LL ring weakly coupled by tunneling contacts to the leads
(for details of the tunneling coupling, see Appendix~\ref{App:Dyson}). We consider a symmetric setup with both point-like contacts having the same tunneling rate
and both arms of the interferometer having the same length. We assume that the Coulomb interaction between electrons on the ring is screened by a ground plane and take the interaction to be pointlike. Throughout the paper we focus on the 
regime of relatively high temperatures
$$E_F \gg T\gg\Delta\gg\Gamma_0~,$$ where $\Gamma_0$ is the tunneling rate in the absence of electron-electron interactions and $E_F$ is the Fermi energy, in which there emerges the new interesting physics, discussed already at the qualitative level in Sec.~\ref{s1}, related to the interplay of SCS and tunneling.
Let us estimate $\Delta$ for realistic systems. For example, for a single-channel GaAs ring
of the radius  $R=L/2\pi=200$ nm, where $L$ is the circumference of the ring, in  the range of  $E_F = 0.02-0.2$\,eV, we find the level spacing at the Fermi level
$\Delta \simeq (0.1-0.4) \times 10^{-2}$\,eV.
 The condition $T>\Delta$ is seen to  be easily satisfied for realistic  experimental conditions.

One of the consequences of taking the high-$T$ limit $T\gg\Delta$ (for $\alpha\alt 1$) is that the classical effect of Coulomb blockade of charge transport through the ring can be neglected. Moreover, in this limit, the number of tunneling channels in the temperature window around the Fermi level is large, of the order of $T/\Delta$. Most importantly in the context of SCS, the condition $T\gg\Delta$ also means that the spin and charge components of the electron propagator inside the ring each have a characteristic spatial extent ($v/T$ and $u/T$, respectively) which is much smaller than the distance between the contacts to the leads. That is, the spin and charge excitations propagate ballistically with velocities $v$ and $u$ and only rarely ``collide'' with the contacts and with each other (see Fig.~\ref{fig1}), in accordance with the picture outlined in Sec.~\ref{s1} \cite{footnote-park15}. This essentially simplifies the two-particle dynamic correlation functions and we use this condition extensively from the very beginning.

For a discussion of the general case of an arbitrary relation between $T$ and the characteristic level spacing in an {\it isolated} finite-length spinful LL, see Ref.~\cite{mattsson97} for a piece of the LL between two hard walls (``quantum dot") and Refs.~\cite{eggert97,pletyukhov06} for a quantum ring made of it. The role of SCS in transport through a 1D quantum dot in the regime of Coulomb blockade was intensively studied in Refs.~\cite{kleimann00,braggio01,cavaliere04,cavaliere04a}. The transport properties of a ring weakly coupled to the leads in the presence of SCS were investigated in the limit $T\ll\Delta$ (and away from the transmission resonances, i.e., in the valleys of Coulomb blockade) in Refs.~\cite{kinaret98,pletyukhov06}. Focusing on either isolated systems or transport in the low-$T$ limit, these studies do not discuss the peculiar interplay between SCS and tunneling that becomes apparent for $T\gg\Delta$ and is the subject of the present work. As already mentioned in Sec.~\ref{s1}, the effect of finite $T\agt\Delta$ on the AB oscillations in a spinful LL cannot be mimicked by averaging \cite{jagla93,hallberg04,meden08} the zero-$T$ transmission coefficient over energy, which misses the interaction-induced decay of single-electron excitations at finite $T$ as well as the ZM dephasing.

We assume that electron-electron backscattering is absent and consider a ring made of a single-channel wire with otherwise generic interactions characterized by four coupling constants: $\alpha_{2\|},\alpha_{2\perp},\alpha_{4\|}$, and $\alpha_{4\perp}$ \cite{giamarchi04}. For the most part, the paper is focused on the study of a symmetric model, isotropic in chirality and spin spaces, with $\alpha_{2\|}=\alpha_{2\perp}=\alpha_{4\|}=\alpha_{4\perp}\equiv\alpha$. Importantly, the isotropic model fully captures the physics of SCS. However, as already mentioned in Sec.~\ref{s1}, the ZM dephasing mechanism, which was shown to be dominant in the spinless case \cite{dmitriev10}, is ineffective in the case of full isotropy. For this case, we obtain an analytical result for the conductance in terms of a phenomenologically introduced dephasing rate $\Gamma_\varphi$ (which may arise from an external bath). By going beyond this model, we demonstrate that a violation of isotropic symmetry leads to two effects: an additional ZM-induced splitting of the AB resonances and the emergence of ZM dephasing. Remarkably, the dephasing action is then described by an equation analogous to that in the spinless case.

\subsection{SCS: Green functions in an infinite system}
\label{s2a}

We start by considering the fully isotropic (in spin and chirality spaces) model introduced above. We write first the corresponding expression for the coordinate-time Green function per spin in the Matsubara representation in an infinite spinful LL \cite{solyom79,voit94},
$${\cal G}(x,\tau)={\cal G}^+ (x,\tau)+{\cal G}^- (x,\tau)~,$$
where $+$ and $-$ denote right- and left-moving fermions, respectively (here and below, $\hbar=1$):
\BEA
&&\hspace{-0.5cm}{\cal G}^{+}(x,\tau)={\cal G}^-(-x,\tau)
= - \frac{i}{2 \pi \sqrt{u v}} \nonumber\\
&&\hspace{-0.5cm}\times\left\{
\frac{\pi T}{\sinh [\pi T(x/v + i\tau)]} \frac{\pi T}{\sinh [\pi T(x/u + i\tau)]}\right\}^{1/2}
\label{18} \\
&&\hspace{-0.5cm}\times\left\{ \frac{\pi T /D}{\sinh [\pi T (x/u + i\tau)]}
\frac{\pi T / D}{\sinh [\pi T(x/u - i\tau)]}\right\}^{\alpha_b/4}. \nonumber
\EEA
Here, $D$ is the ultraviolet cutoff in energy space,
\be
\alpha_b=(u-v)^2/2uv~,\quad u=v(1+4\alpha)^{1/2}~,
\ee
and
\be
\alpha=V_0/2\pi v~,
\label{4}
\ee
with $V_0$ being the zero-momentum Fourier component of the interaction potential.

The main (``chiral" or ``square-root") approximation below, which we make following Ref.~\cite{jagla93} (but, in contrast to it, not focusing on the case of ``commensurate" $v/u$ given by a simple fraction) and Ref.~\cite{yashenkin08}, is to keep, in the electron self-energy, only the terms of leading (linear) order in $\alpha\ll 1$. That is, we retain the difference between $u$ and $v$,
\be
u \simeq (1+2\alpha)v~,
\label{uv}
\ee
and put $\alpha_b\sim {\cal O}(\alpha^2)$ to zero in Eq.~(\ref{18}). Note that it is the appearance of the two velocities for each chirality in the
single-particle correlator ${\cal G}^{\pm}$ that signifies SCS. Within this approximation, the right and left electrons are chiral, and the charge and spin velocities enter the electron Green function for each chirality in a symmetric way. From the point of view of symmetry in chirality space, the approximation becomes exact when only interactions between electrons of the same chirality are kept (see the discussion of the generic model in Sec.~\ref{sec:deph} below). It is worth noting that the consistency of the square-root approximation within the isotropic LL model requires that the difference of the two velocities is small [Eq.~(\ref{uv})].

Relying on the chiral approximation, which fully captures SCS, allows us to obtain closed analytical expressions for the conductance and the dephasing rate without sacrificing anything of importance as far as the essence of the effect of SCS on the AB oscillations is concerned. Importantly, the neglected higher-order effects, which produce asymmetry between the spin and charge in Eq.~(\ref{18}), correspond to ``fractionalization" \cite{pham00,lehur08,steinberg08,karzig11,calzona15, calzona16,acciai17,brasseur17}  of the charge sector in much the same way as in the spinless case \cite{dmitriev10}, and likewise do not lead to a suppression of the interference. Indeed, the fractionalized charges emitted at one contact, moving in the opposite directions with the \textit{same velocity} $u$, collide every time they pass by the contacts. They sum up to the ``unfractionalized" electron charge at the contact, so that charge fractionalization does not manifests itself in the ring geometry \cite{remark1}. This should be contrasted with spin-charge separated excitations (plasmons and spinons) which move in the same direction with \textit{different velocities} and, therefore, pass by a contact at the same time only rarely. The dynamical properties of charge fractionalization on the one hand and those of SCS on the other are thus essentially different.

In the real-time representation, the retarded Green functions \cite{gornyi07} corresponding to the square-root approximation read
\begin{align}
&{\cal G}^+_R(x,t)={\cal G}_R^-(-x,t) \simeq \frac{i\theta(t)}{ \pi  \sqrt{u v}}
\label{Green}
\\ \nonumber
& \times \! {\rm Im}\,  \frac{ \pi T} { \sqrt{\sinh [ \pi T (x/u-t +i0) ] \sinh [ \pi T (x/v-t +i0) ]}}
\end{align}
(in the energy-momentum representation, the expressions for the spin-charge separated Green functions for $\alpha\ll 1$ can be found in Ref.~\cite{yashenkin08}). Note that, within the square-root approximation, the time dependence of the Green functions ${\cal G}^\pm_R(x,t)$ at $x=0$, which determines the tunneling density of states, coincides with that for the noninteracting Green functions.

\subsection{SCS: Isolated ring}
\label{subsec:IIB}

Let us now discuss the SCS effects in a single-channel quantum ring of length $L$ threaded with the magnetic flux $\phi$. We first consider an isolated (not coupled to the leads) ring. In the finite-size system, in addition to plasmons and spinons, one should take into account homogeneous ZM excitations. Such excitations are characterized by eigenenergies $E_{\rm ZM}$ that are determined by the total numbers $N_\pm^{s}$ of right- and left-moving ($\pm$) particles with a given spin projection $s=\uparrow,\downarrow$ and the chemical potential $\mu$ (the same for all sorts of particles) fixed by the leads. Under the assumption of  full isotropy of interactions in chirality and spin spaces within the ring,
the ZM energy  is given by
\cite{pletyukhov06,meden08}:
\BEA
E_{\rm ZM}&=&\frac{\Delta_u K_\rho}{8}\left[\,\frac{1}{K^2_\rho}\left(N_c-4N_0\right)^2+\left(J_c-4\phi\right)^2
\nonumber
\right.
\\
&+& \left. 2\left(N_{+}^{\uparrow}-N_{+}^{\downarrow}\right)^2+
2\left(N_{-}^{\uparrow}-N_{-}^{\downarrow}\right)^2\,\right]~,
\EEA
where
\BEA
N_c&=&N_{+}^{\uparrow}+N_{+}^{\downarrow}+N_{-}^{\uparrow}+N_{-}^{\downarrow}~, \label{N_c}\\
J_c&=&N_{+}^{\uparrow}+N_{+}^{\downarrow}-N_{-}^{\uparrow}-N_{-}^{\downarrow}~, \label{J_c}
\EEA
$\Delta_u=2\pi u/L$, and $K_\rho=v/u\simeq 1-2\alpha$ is the Luttinger constant (for the charge sector). The current and spin contributions to $E_{\rm ZM}$ are characterized by the ``noninteracting'' spacing $\Delta=\Delta_u K_\rho=2\pi v/L$. It is worth noting that the ZM excitations are not factorizable into independent charge and spin parts. The total number of electrons in the ring is controlled by $N_0$ (which, in turn, is determined by $\mu$).

The equilibrium value of an observable ${O}$ is averaged over ZM fluctuations (in a grand canonical ensemble of isolated rings) according to
\be
\langle {O} \rangle_{\rm ZM}\equiv \frac{ \sum\limits_{N} {O}_N e^{-E_{\rm ZM}(N)/T}} { \sum\limits_{N} e^{-E_{\rm ZM}(N)/T}}~,
\label{OZM}
\ee
where
\be
{O}_N= {O}_{N}^{\uparrow} + {O}_{N}^{\downarrow}
\ee
is the sum of the observable for spin-up and spin-down electrons and a given set of $N=(N_{+}^{\uparrow},N_{+}^{\downarrow},N_{-}^{\uparrow},N_{-}^{\downarrow})$. We do not discuss here the spin-orbit and Zeeman couplings, so that spin-rotational symmetry is preserved: ${O}_N^{\uparrow}={O}_N^{\downarrow}$. Therefore, below we omit the spin index.

For a given spin projection, the retarded Green function of right movers ${\cal G}^{+}_{R}(x,t)$ is given, in the closed ring, by a product
\be
{\cal G}^{+}_{R}(x,t)=\mathcal{G}_{\rm ZM}^{+}(x,t)\ \mathcal{G}^{+}_{\rm SC}(x,t)
\label{Green-finite}
\ee
of the ZM factor $\mathcal{G}_{\rm ZM}^{+}(x,t)$ and the factor $\mathcal{G}^+_{\rm SC}(x,t)$ which describes excitations with nonzero momenta and, in turn, factorizes into a product of the spin and charge parts (hence ``SC''), similarly to Eq.~(\ref{Green}). For given $N_+$, the ZM factor in Eq.~(\ref{Green-finite}) is written as
\begin{equation}
\mathcal{G}_{\rm ZM}^{+}(x,t)=\exp[-\,i \delta E^+_{\rm ZM} t  + 2\pi i x(N_{+}+1)/L \,]~,
\label{GZM}
\end{equation}
where
\begin{equation}
\delta E^{+}_{\rm ZM}= E_{\rm ZM}(N_{+}+1)-E_{\rm ZM}(N_{+})
\label{dEZM}
\end{equation}
is the variation of the ZM energy with changing $N_+$ by unity. The retarded Green function for left movers $\mathcal{G}^-_{R}(x,t)$ is obtainable from $\mathcal{G}^+_{R}(x,t)$ by changing $x\to -x$ and $N_+\to N_-$.

Within the chiral approximation (\ref{Green}), the factor $\mathcal{G}^{+}_{\rm SC}$ reads
\begin{align}
&\mathcal{G}^{+}_{\rm SC}(x,t)\!\simeq\!\frac{i\theta(t)}{\pi\sqrt{u v}} ~{\rm Im} \sqrt{\,\sum_{nm}A_u(x_n,t-i0)A_v(x_m,t-i0)},
\notag
\\
\label{g}
\end{align}
where
\be
A_u(x,t)=\dfrac{\pi T}{\sinh\left[\pi T\left(x/u-t\right)\right]}~,
\label{Au}
\ee
$A_v(x,t)$ is given by the same expression with $u$ substituted with $v$. The argument
$$x_n=x+nL$$
in $A_{u,v}(x_n,t)$ corresponds to the paths with multiple revolutions around the ring and the summation is taken over all integer $n$ and $m$. Equation~\eqref{g} is obtained from Eq.~\eqref{Green} by ``replicating" the spin and charge factors, namely by replacing $x \to x_n$ in the plasmon factor and $x \to x_m$ in the spinon one, and summing over $n$ and $m$. Importantly, the spin and charge factors are replicated independently---this follows most directly from the bosonization approach, where each of these factors is determined by an independent bosonic field, spinon or plasmon, which is periodic in real space with the period $L$. In this way, we obtain a double sum over $n$ and $m$ under the square root sign in Eq.~\eqref{g}. Note that Eq.~\eqref{Green-finite} with $\mathcal{G}^{+}_{\rm SC}(x,t)$ from Eq.~(\ref{g}) reproduces the Green function of a noninteracting closed ring by equating $u=v$ and using the property $${\rm Im}\,A_v(x,t-i0)=-\pi\delta(t-x/v)~.$$

In the limit $T\gg\Delta$, the calculations can be  simplified by noticing that the function $A_u(x,t)$ [and similarly $A_v(x,t)$] is then sharply peaked in time and space, within the small time and space intervals of width $\delta t\sim 1/T$ and $\delta x\sim u/T$, respectively. Namely,  $\delta t\ll 1/\Delta$ and $\delta x\ll L$. This, in turn, means that the peaks  associated with different terms in the sum over $n$ in Eq.~(\ref{g}) are well separated in space and time, and similarly for the sum over $m$. This allows us to commute the square root of the sum in Eq.~(\ref{g}) into a sum of square roots,
$$\sqrt{\sum_{nm}(\dots)}\to \sum_{nm}\sqrt{(\dots)}~,$$
and write $\mathcal{G}^{+}_{\rm SC}(x,t)$ as  \cite{fractionalization}
\begin{align}
&\mathcal{G}^{+}_{\rm SC}(x,t)\simeq\frac{\theta(t)}{2\pi\sqrt{u v}}
\notag
\\
&\times \sum\limits_{nm} \sum \limits_{\eta=\pm1}\eta \sqrt{\,A_u(x_n,t -i\eta\,0)A_v(x_m,t-i\eta\,0)}~.
\label{ga}
\end{align}
Equations (\ref{Green-finite}), (\ref{GZM}), and (\ref{ga}) define the spin-charge separated electron Green function in a closed ring for $\alpha\ll 1$ and $T\gg\Delta$.  An important property of the spin-charge factor in the Green functions of right and left movers, which we use below, is
\be
\mathcal{G}^{+}_{\rm SC}(L/2,t)=\mathcal{G}^{-}_{\rm SC}(L/2,t)~.
\label{L2}
\ee
The time Fourier transform of the functions $\mathcal{G}_R^\pm(x,t)$ defines the Green functions $G_R^\pm(\epsilon,x )$ in the energy-space
representation. The explicit form of $G_R^\pm(\epsilon,x )$ is derived in Appendix \ref{properties}.

\subsection{SCS: Tunneling conductance of a ring}
\label{s2c}

Now we include the tunnel coupling between the ring and the leads (Fig.~\ref{fig1}). Both the ZM and plasmon-spinon SCS factors are modified by the ``opening" of the ring. Before we proceed to a discussion of these modifications (this will be done in Sec.~\ref{s3b}), let us see in what combination the ZM and SCS factors enter the conductance through the ring. In what follows, we assume for simplicity that the Fermi velocity in the leads $v_F$ is equal to the velocity of excitations in the noninteracting ring: $v_F=v$.

We formalize our approach to transport of interacting electrons through the ring tunnel-coupled to two Fermi reservoirs in terms of the Kubo formula, in which the process of single-electron tunneling across the ring is accompanied by tunneling of other electrons which serve as a dephasing environment for the tunneling electron. In the noninteracting limit, the conductance ${\rm G}(\phi)$ is written as the energy-averaged single-electron transmission coefficient $\cal T(\phi)$:
\be
{\rm G}(\phi)=2\times\frac{e^2}{2\pi}\,{\cal T}(\phi)=\frac{e^2}{\pi}\,\langle{\rm T}(\phi,\epsilon)\rangle_T~,
\label{1} \ee
where ${\rm T}(\phi,\epsilon)$ is the transmission coefficient at energy $\epsilon$, the factor of 2 accounts for spin, and
\be
\langle\cdots\rangle_T=\int\!d\epsilon\,(-\partial_\epsilon f)(\cdots)
\label{T-aver}
\ee
denotes the thermal averaging over $\epsilon$ with the Fermi distribution function, $f=\{1+\exp[(\epsilon-\mu)/T]\}^{-1}$, characterized by the chemical potential $\mu$ in the leads.

Importantly, as was shown in Ref.~\cite{dmitriev10} (see also Appendix~\ref{App:Dyson}), in the vicinity of $\phi=1/2$, backscattering by the tunnel contacts can be neglected (both in the noninteracting and interacting spinless cases). Within the square-root approximation of Sec.~\ref{s2a}, this is also true for spinful interacting electrons. As a result, the right and left sectors of our model are decoupled (for given $N_\pm$). This allows us to use Eq.~\eqref{1} (averaged over ZM fluctuations) in which SCS modifies the energy dependence of ${\rm T}(\phi,\epsilon)$.

Since we assumed the arms of the interferometer to be of equal length (with the coordinates of the point-like contacts $x=0$ and $x=L/2$), the transmission coefficient is expressible in terms of the retarded Green function in the energy-coordinate representation as
\begin{align}
&&\hspace{-0.4cm}{\rm T}(\phi,\epsilon) \! =\!
v^2|t_{\rm tun}|^2|t_{\rm out}|^2  \!\left \langle\!  \left | G_{R}^+(\epsilon,L/2)\!+\!G_{R}^-(\epsilon,L/2) \right|^2 \right \rangle_{\!\rm ZM}\! ,
\nonumber
\\
&&\label{15}
\end{align}
where we used the short-hand notation $G_R^\pm(\epsilon,L/2)$ for the Green function $G_R^\pm(\epsilon,0,L/2)$ that connects the contacts and takes into account tunneling to the leads. In Eq.~\eqref{15}, electron-electron interactions are accounted for within the square-root approximation for the Green functions forming the fermion loop for the density-density response function and through the ZM averaging. The dimensionless bare (noninteracting) amplitudes $t_{\rm tun} $ and $t_{\rm out}$ in Eq.~\eqref{15} describe tunneling of an electron to and out of the ring, respectively (see Appendix \ref{App:Dyson}) \cite{footnote-contacts}. If, as assumed, the contacts are point-like and right-left symmetric (which together means time-reversal symmetry for scattering at the contact),
$$t_{\rm tun}=t_{\rm out},$$
independently of the strength of tunneling. The tunneling rate for the noninteracting ring is  given by \cite{dmitriev10}:
$$\Gamma_0 \simeq \frac{ |t_{\rm tun}|^2}{\pi } \Delta~.$$

To make use of the results obtained in Sec.~\ref{subsec:IIB} for an isolated ring, it is convenient to transform the thermal averaging (\ref{T-aver}) in Eq.~(\ref{1}) with ${\rm T}(\phi,\epsilon)$ from Eq.~(\ref{15}) into the real-time representation, which leads to
\begin{align}
&{\cal T(\phi)}=\frac{v}{u}\left(\frac{\pi \Gamma_0}{\Delta}\right)^2
\label{19a}
\\
&\times
\!\int_0^\infty\!\!dt\!\int_{-t}^\infty\!\!d\tau\,\frac{\pi T\tau}{\sinh\pi T\tau}
 g(t+\tau)g^*(t)F_{\rm ZM}(t,\tau,\phi)~.
\notag
\end{align}
Here, the function $g(t)$ describes propagation of an electron from one contact to the other in time $t$, including multiple revolutions around the ring and tunneling at the contacts. The function $F_{\rm ZM}(t,\tau,\phi)$ stems from the averaging of the fermionic loop over the ZM fluctuations. The factorization of the integrand in Eq.~\eqref{19a} relies on the ``right-left'' symmetry of the model (equal arms and chirality separation). Importantly, in the Green function $G_R^\pm(\epsilon,L/2)$ transformed into the coordinate-time representation, the functions $\mathcal{G}_{\rm ZM}^{\pm}(L/2,t)$ and $g(t)$ factorize, similarly to Eq.~\eqref{Green-finite}, even in the presence of tunneling, as can be seen from the expressions derived in Appendix~\ref{properties}. We stress that, because of the ``right-left'' symmetry, the factor $g(t)$ is the same for both right and left movers [for an isolated ring, this follows explicitly from Eq.~\eqref{L2}].

Because of tunneling, $g(t)$ decays with time, as a result of which the $t$ integral in Eq.~(\ref{19a}) converges at $t\to\infty$. It is instructive, however, to first examine the behavior of the integrand in a closed ring (for more details, see Sec.~\ref{s3a}). In this case, we have
\be
g(t)= (u v)^{1/2}\mathcal{G}_{\rm SC}^+(L/2,t)~,
\label{22}
\ee
where $\mathcal{G}^+_{\rm SC}(x,t)$ is given by Eq.~(\ref{ga}). The factor $F_{\rm ZM}(t,\tau,\phi)$ is expressed through the ZM functions from Eq.~\eqref{GZM} as
\begin{align}
&F_{\rm ZM}(t,\tau,\phi)=
\left\langle [\mathcal{G}_{\rm ZM}^{+}(L/2,t)+\mathcal{G}_{\rm ZM}^{-}(L/2,t)]\right.
\label{FZG} \\
&\qquad \quad \times \left.[\mathcal{G}_{\rm ZM}^{+}(L/2,t+\tau)
+\mathcal{G}_{\rm ZM}^{-}(L/2,t+\tau)]\right\rangle_{\!ZM}.
\notag
\end{align}
The next step is to use the parameter $T/\Delta \gg 1$ to simplify $F_{\rm ZM}(t,\tau,\phi)$. Because of the $1/\sinh (\pi T\tau)$ factor in Eq.~(\ref{19a}), we have $|\tau|\alt 1/T\ll 1/\Delta$. This, for typical deviations of $N_\pm$ from $N_0$, allows us to neglect the dependence on $\tau$ in $F_{\rm ZM}(t,\tau,\phi)$, yielding
\begin{align}
&F_{\rm ZM}(t,0,\phi)=2{\rm Re}\left[1 + K_{\rm ZM}(t)e^{-2i\Delta \phi t}\right]~,
\label{FZM}
\end{align}
where
\begin{align}
K_{\rm ZM}(t)
&=\left\langle e^{i\Delta(N_+-N_-)t+i\pi(N_+-N_-) }\right\rangle_{\rm ZM}~.
\label{KZM}
\end{align}
Here, we used
\be
\delta E^{+}_{\rm ZM}-\delta E^{-}_{\rm ZM} =  \Delta (N_+ -N_-  - 2 \phi)~,
\label{24}
\ee
which follows directly from the definition \eqref{dEZM} of $\delta E^{\pm}_{\rm ZM}$. After setting $\tau=0$ in $F_{\rm ZM}(t,\tau,\phi)$, the SCS dynamics in the ring decouples from the ZM dynamics at any point in time and is then encoded in Eq.~(\ref{19a}) through the function
\be
K(t)=g^*(t)\int_{-t}^{\infty}\!\!d\tau\,\frac{\pi T\tau}{\sinh\pi T\tau}\,g(t+\tau)~.
\label{12}
\ee

In a tunnel-coupled ring, both the functions $K(t)$ and $K_{\rm ZM}(t)$ are modified. First, $K(t)$ is still represented in terms of the single-electron propagator $g(t)$ in exactly the same way as in Eq.~(\ref{12}), but $g(t)$ is now dressed by tunneling vertices and describes the SCS dynamics with the inclusion of tunneling-induced decay. Second, $K_{\rm ZM}(t)$ in the tunnel-coupled ring is given by the average of the same exponential function as in Eq.~(\ref{KZM}) but the meaning of the averaging is different. Specifically, $N_\pm$ become functions of $t$, because of the exchange of electrons between the ring and the leads, so that the averaging goes over the equilibrium dynamic fluctuations in the open system rather than over the grand canonical ensemble in which $N_\pm$ were $t$-independent numbers.

The transmission coefficient (\ref{19a}) at $T\gg \Delta$ can thus be written as
\be
{\cal T(\phi)}\!=\!\frac{2v}{u}\left(\frac{\pi \Gamma_0}{\Delta}\right)^2
\! {\rm Re} \int_0^\infty\! \!dt\, K(t)\,\left[1+K_{\rm ZM}(t)e^{-2i\Delta \phi t}\right]~.
\label{25}
\ee
Further, it can be represented as a sum of the classical (${\cal T}_c$, independent of $\phi$) and quantum (${\cal T}_q$, describing the AB interference)
contributions:
\be
{\cal T}(\phi)={\cal T}_c +{\cal T}_q(\phi)~,
\ee
where ${\cal T}_c$ and ${\cal T}_q$ correspond to the first and second terms in the square brackets in Eq.~(\ref{25}), respectively. Note that {\it both} contributions are affected by SCS, whose dynamics is encoded in the factor $K(t)$.

\section{Dynamics of SCS}
\label{s3}

The dynamical properties of SCS (see, e.g., Refs.~\cite{calzona15,calzona16,acciai17,hashisaka17}) are described, in the ring geometry, by the function $K(t)$. As mentioned in Sec.~\ref{s2c}, we first examine the behavior of $K(t)$ in a closed ring in Sec.~\ref{s3a}. The SCS dynamics in the presence of tunneling will be analyzed in Sec.~\ref{s3b}.

\subsection{SCS dynamics: Isolated ring}
\label{s3a}

We start with $K(t)$ in a closed ring. From  Eqs.~\eqref{ga} and (\ref{22}), we have
\begin{align}
g(t)&=\frac{i}{\pi}\sum\limits_{nm}\,{\rm Im}\,\sqrt{\,A_u(L_n,t-i0)A_v(L_m,t-i0)}
\label{g1}
\\
&=\frac{1}{i\pi} \sum \limits_{nm}\Theta_{nm}(t)\sqrt{ | A_u(L_n,t)A_v(L_m,t) |}~,
\label{g-theta}
\end{align}
defined for $t>0$, where
\be
L_n=(n+1/2)L~
\label{12b}
\ee
and
$$\Theta_{nm}(t)=\theta \left[\left(L_n/u-t\right)\left(t-L_m/v\right)\right]$$ with $\theta(t)$ being a step function. Each of the terms in the function $g(t)$ is only nonzero within the time interval between $L_n/u$ and $L_m/v$, has square-root singularities at the end points, and is suppressed if the width of the interval
\be
(\Delta t)_{nm}=|L_n/u-L_m/v|
\label{12d}
\ee
is much larger than $1/T$ as $\exp [-\pi T(\Delta t)_{nm}/2]$. Note that $g(t)\simeq {\rm const}(t)$ inside the above interval in the limit $T(\Delta t)_{nm}\gg 1$ except for the narrow regions of width of the order of $1/T$ in the vicinity of the end points.

A useful way of representing $K(t)$ in terms of $g(t)$ from Eq.~(\ref{g1}) is
\be
K(t)=-\frac{1}{4\pi^2}\!\sum\limits_{\eta\eta'}\eta\eta'\!\!\int_{-t}^{\infty}\!\!\!d\tau\,\frac{\pi T\tau}{\sinh\pi T\tau}\,w_c^{\eta \eta'}(t,\tau)w_s^{\eta \eta'}(t,\tau)~,
\label{12a}
\ee
where $\eta,\eta'=\pm1$ and
\begin{align}
&w_c^{\eta\eta'}(t,\tau)\!\simeq\!\sum\limits_n\!\!\sqrt{A_u(L_n,t-i\eta\,0)A_u(L_n,t+\tau-i\eta'0)}~,
\label{rho-c}\\
&w_s^{\eta\eta'}(t,\tau)\!\simeq\!\sum\limits_n\!\!\sqrt{A_v(L_n,t-i\eta\, 0)A_v(L_n,t+\tau -i\eta'0)}
\label{rho-s}
\end{align}
are the elements of the charge $(c)$ and spin $(s)$ single-particle ``fusion" matrices in $\eta$ space at $x=L/2$. Note that each of them contains, upon substitution of Eq.~(\ref{g1}) in Eq.~(\ref{12}), a double sum over windings; however, in the limit $T\gg\Delta$, the contribution of nondiagonal terms in the double sums is small and can be neglected, which is done in Eqs.~(\ref{rho-c}) and (\ref{rho-s}). This is because of the defining property of $w_c^{\eta\eta'}(t,\tau)$ and $w_s^{\eta\eta'}(t,\tau)$, which is that each of them is characterized by only one velocity: $u$ in the former case and $v$ in the latter. The SCS dynamics can now be visualized by straightforwardly generalizing the definition of $w_c^{\eta\eta'}(t,\tau)$ and $w_s^{\eta\eta'}(t,\tau)$ to arbitrary $x$ and thinking of the charge and spin ``wave packets" which rotate independently around the ring with velocities $u$ and $v$, respectively.

\begin{figure}[ht!]
\leavevmode \epsfxsize=8cm
\centering{\epsfbox{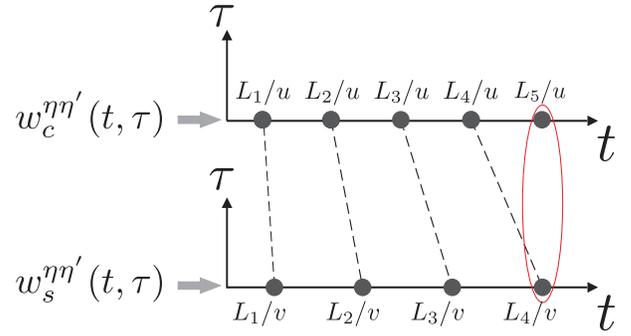}}
\caption{Illustration of the SCS dynamics. Black dots: Regions in $(t,\tau)$ space where the charge and spin densities are peaked. The characteristic size of these regions (the size of the dots) is of the order of $1/T$ along both the $t$ and $\tau$ axes. In this particular example, a spin-charge collision (marked by the red oval) happens at $t\simeq L_5/u\simeq L_4/v$, with $L_n$ from Eq.~(\ref{12b}).}
\label{fig2}
\end{figure}

The complex functions $w_c^{\eta\eta'}(t,\tau)$ and $w_s^{\eta\eta'}(t,\tau)$ are periodic in $t$ and concentrated around  $t=L_n/u$ and $t=L_n/v$ with $n=0,1,2,\ldots$, respectively. Each of the peaks  has  a characteristic width of the order of $1/T$ in both the $\tau$ and $t$ direction. This is schematically depicted in Fig.~\ref{fig2}, where the regions in $(t,\tau)$ space in which the charge and spin wave packets are peaked at $x=L/2$ are marked by black dots. Now, note that there exist such winding numbers for the charge ($n$) and spin ($m$) that the charge and spin wave packets are peaked at $x=L/2$ {\it simultaneously}. Namely, the spin-charge collision occurs for $n$ and $m$ obeying
\be
(\Delta t)_{nm}\lesssim 1/T
\label{26}
\ee
(in Fig.~\ref{fig2}, the collision is shown for $n=5$ and $m=4$).

While correct in spirit, the above picture of spin-charge collisions is not quite right in one important aspect: it does not yet include the destructive interference between the wave packets, i.e., the cancellation between the different contributions to $K(t)$ after the summation over $\eta$ and $\eta'$. For small $\alpha$, the cancellation tends to make the widths of the peaks in $K(t)$ shorter compared to $1/T$. Indeed, in the strictly noninteracting limit, the spin and charge move together, Eq.~(\ref{26}) is always satisfied, and $K(t)$ is a strictly periodic series of $\delta$ functions with a period $2\pi/\Delta$; specifically,
\be
K(t)=\sum_n\delta(t-L_n/v)
\label{27}
\ee
in the limit of $T\gg\Delta$. For $\alpha=0$, the width of the peaks in $K(t)$ is thus exactly zero, although the wave packets in Eqs.~(\ref{rho-c}) and (\ref{rho-s}) have a finite width at $1/T\neq 0$. For $\alpha\neq 0$, we have, after summing over $\eta$ and $\eta'$:
\BEA
K(t)\!&=&\frac{1}{\pi^2}\!\int_{-t}^{\infty}\!\!d\tau\,\frac{\pi T\tau}{\sinh\pi T\tau} \sum_{nm} C_n(t,\tau)S_m(t,\tau)\nonumber \\
&\times &\!
\Theta_{nm}(t) ~\Theta_{nm}(t+\tau)
\label{28}
\EEA
where
\BEA
&&\hspace{-1cm}C_n(t,\tau)\nonumber\\
&&\hspace{-1cm}=\frac{\pi T}{\sqrt{\sinh |\pi T(L_n/u-t)|\sinh |\pi T (L_n/u-t-\tau)|}}~,
\label{29a}\\[0.3cm]
&&\hspace{-1cm}S_n(t,\tau)\nonumber\\
&&\hspace{-1cm}=\frac{\pi T}{\sqrt{\sinh |\pi T(L_n/v-t)|\sinh |\pi T (L_n/v-t-\tau)|}}~.
\label{29b}
\EEA
The factor $\Theta_{nm}(t)\Theta_{nm}(t+\tau)$ restricts both $t$ and $|\tau|$  to within the interval between $L_n/u$ and $L_m/v$.

In order to compare Eq.~\eqref{28} with the noninteracting result, Eq.~\eqref{27},  it is instructive to rewrite the sum over $n$ and $m$ as a sum over $n$ and the difference between  the charge and spin winding numbers $k=n-m$. Each term in the sum over $n$ for given $k$ yields a resonant contribution to $K(t)$ similar to the   delta-function peaks in Eq.~\eqref{27}. One of the modifications brought about by SCS is the broadening of these peaks, as can be seen from the second line of Eq.~(\ref{28}). Namely, the time interval between the exact borders of the ``resonance" in $K(t)$ that occurs after $n$ revolutions of the charge wave packets and $n-k$ revolutions of the spin ones is rewritten as
\be
(\Delta t)_{n,n-k}= \left|\,k-\left(1-\frac{v}{u}\right)\left(n+\frac{1}{2}\right)\right|\frac{2\pi}{\Delta}~.
\label{30}
\ee

A more dramatic effect of SCS on the peaks in $K(t)$ is a deep periodic modulation of the amplitude of the envelope of the series of peaks, which is a remarkable consequence of the condition (\ref{26}). These peaks are enumerated by the integer number $k$. For given $k,$ the equation $(\Delta t)_{n,n-k}=0$ (``exact spin-charge collision") has a solution $n=n_k$ (assuming for a moment that $n$ is a continuous variable), where
\be
n_k+\frac12= k\frac{u}{u-v}~.
\label{31}
\ee
For integer charge winding numbers $n$ close to $n_k$, the (generically nonzero) interval $(\Delta t)_{n,n-k}$ is small in the sense of the condition (\ref{26}), provided $\alpha$ is sufficiently small. Specifically, for $\alpha\ll\Delta/T$, the characteristic number of peaks around $n_k$ that are not suppressed by the temperature is given by $\delta n\sim\Delta/\alpha T$. In the limit of $T\gg\Delta$, we have $\delta n\ll n_{k+1}-n_k$, so that the bunches of peaks centered at $n \simeq n_k$ are well separated from each other.

With these ingredients, the picture that emerges in the time domain is that of a train of narrow ``double-horn'' peaks of width
\be
(\Delta t)_{n,n-k} \simeq \frac{4\pi}{\Delta}\alpha|n-n_k|~,
\label{33}
\ee
with
\be
n_k \simeq \frac{k}{2\alpha} +k - \frac{1}{2}, \qquad  k=0,1,2\ldots,
\label{nk}
\ee
and the nearest-neighbor spacing $2\pi/\Delta$, which are grouped together in bunches of width of the order of $1/\alpha T$ centered at
\be
t_k=\pi k/\alpha\Delta~.
\label{34}
\ee
As we will see below, the subleading term $k-1/2$ should be kept in Eq.~\eqref{nk} for the calculation of $K_{\rm ZM}(t)$.

\begin{figure}[ht!]
\leavevmode \epsfxsize=8.5cm
\leavevmode \epsfxsize=8.5cm
\centering{\epsfbox{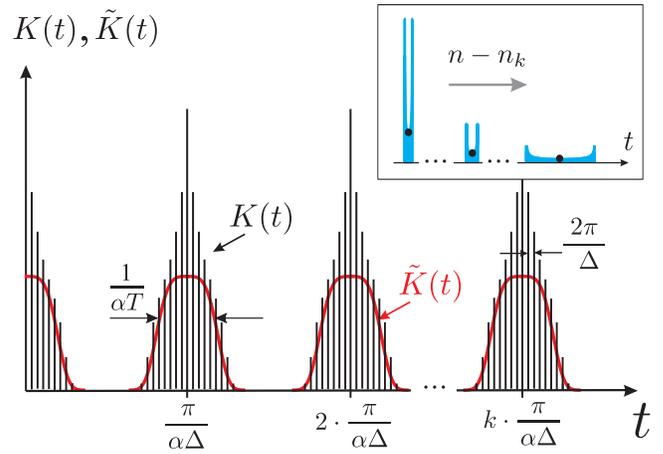}}
\caption{Behavior of the function $K(t)$, which is given by a sequence of narrow peaks (schematically shown by black  vertical lines) grouped into periodic bunches, and its smooth envelope $\tilde K(t)$ (red) in a closed ring. The peaks have a nontrivial ``double-horn'' structure with sharp borders and square-root singularities at these borders as schematically illustrated in the inset. The  parameters of these  peaks evolve with increasing $n-n_k$ (the deviation from the center of the bunch). The height of each of the black vertical lines corresponds to the height of the center of a peak, as shown by black dots in the inset.}
\label{fig3}
\end{figure}

The above picture, illustrated in Fig.~\ref{fig3}, is in accord with the two time scales $\tau_{\rm d}$ and $\tau_{\rm sc}$ introduced in Sec.~\ref{s1}. Specifically, the characteristic bunch width has the meaning of the dwelling time $\tau_{\rm d}$, during which the spin and charge ``stick" to each other:
\be
\tau_{\rm d}=\pi/\alpha T~.
\label{35}
\ee
The distance between the bunches has the meaning of the time between consecutive spin-charge collisions
\be
\tau_{\rm sc}=\pi/\alpha\Delta~.
\label{36}
\ee
Note that, as $|n-n_k|$ is increased, the width of the peaks (\ref{33}) becomes of the order of $1/T$ at the half-height of the bunches. That is, except for the very center of a bunch, the characteristic width of the peaks in $K(t)$ is given by $1/T$ (similar, in this sense, to the peaks in the spin and charge wave packets, illustrated in the cartoon of Fig.~\ref{fig2}).

For $\Delta\tau_{\rm d}\gg 1$, it is useful to introduce the (dimensionless) envelope function $K_0(t)$ for a single bunch of peaks by replacing the double-horn peaks with the delta functions and writing $K(t)$ in the form
\begin{align}
K(t)&\to\sum\limits_{kn}\delta\left[\,t-t_k-\frac{2\pi}{\bar \Delta}(n-n_k)\,\right]K_0(t-t_k)
\label{K18}
\\
&\simeq \sum\limits_{kn}\delta \left[t-  \frac{2\pi}{\bar \Delta}\left(n+\frac12 \right) +\frac{\pi}{\Delta }k \right]K_0(t-t_k)~.
\label{K18b}
\end{align}
The distance $2\pi/\bar\Delta$ between the neighboring peaks of the same bunch in Eqs.~(\ref{K18}) and (\ref{K18b}) is given by
\be
\frac{2\pi}{\bar\Delta}\equiv\pi\left(\frac{1}{\Delta}+\frac{1}{\Delta_u}\right)\simeq\frac{2\pi}{\Delta}(1-\alpha)
\label{Delta-bar}
\ee
(recall that the double-horn peaks are centered in the middle of the interval between $L_n/u$ and $L_{n-k}/v$, see the inset in Fig.~\ref{fig3}). The function $K_0(t-t_k)$ has the meaning of the integral $\int\!dt\,K(t)$ over the period $2\pi/\bar\Delta$ around the point $t$ which belongs to the $k$-th bunch. Using Eq.~\eqref{g-theta}, we obtain $K_0(t-t_k)$ in the scaling form $$K_0(t-t_k)=I[2\pi\alpha T(t-t_k)]~,$$ where the shape of the envelope is given by
\BEA
\phantom{a}\hspace{-1cm}I(z)\!&=&\!\frac{1}{\pi^2}\int_0^z\!dx\int_0^z\!dy\frac{x-y}{\sinh(x-y)}\nonumber\\
&\times&\!\frac{1}{\sqrt{\sinh x\,\sinh (z-x)\,\sinh y\,\sinh(z-y)}}~.
\label{37}
\EEA
Doing the integrals (see Appendix \ref{App:C}), we have
\be
I(z)=\frac{z}{\sinh  z}~.
\label{37a}
\ee
The envelope of the series of bunches $\tilde{K}(t)$ is written as the sum
\be
\tilde{K}(t)=\sum_kK_0(t-t_k)~.
\label{KK0}
\ee

The picture of the densely packed bunches of peaks describes the case of $\alpha\ll\Delta/T$. In the opposite limit, the width of the bunches becomes smaller than the interpeak distance. Then, a given bunch  contains at most one peak  or, typically, no peaks at all, depending on the commensurability between $u$ and $v$.
We leave the discussion of this case and the related commensurability problem  out of the scope of the present paper.
For the results in the literature on the role of the commensurability in transport through the interacting ring see, e.g., Refs.~\cite{jagla93,hallberg04,meden08,rincon09}.

In the above, we have analyzed the behavior of the function $K(t)$ [Eq.~(\ref{12})], which describes the dynamics of SCS, in a closed ring.
Now we turn to the function $K_{\rm ZM}(t)$ [Eq.~(\ref{KZM})] which encodes the ZM dynamics. Consider behavior of $K_{\rm ZM}(t)$ in the vicinity of the $k$-th bunch. Taking $K_{\rm ZM}(t)$ at the times prescribed by the delta functions in Eq.~\eqref{K18b} and
neglecting the ${\cal O}(\alpha^2)$ terms in the exponent, we get
\be
K_{\rm ZM}(t)\simeq  \left \langle e^{-i\alpha\Delta (N_+-N_-) (t-t_k) } \right\rangle_{\rm ZM}~.
\label{KZM1}
\ee
Note that the exponent in Eq.~(\ref{KZM1}) is now explicitly proportional to $\alpha$, in contrast to Eq.~(\ref{KZM}).

\begin{figure}[ht!]
\leavevmode \epsfxsize=8.5cm
\centering{\epsfbox{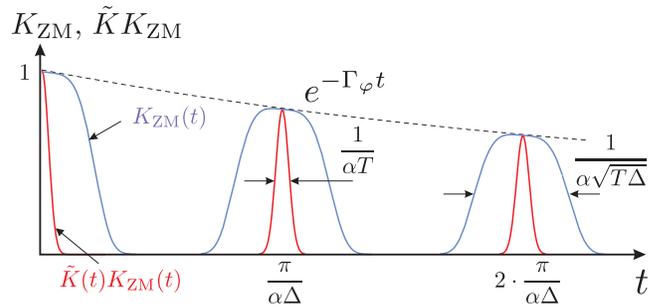}}
\caption{Schematic time dependence of the ZM factor $K_{\rm ZM}(t)$, superimposed onto that of the SCS factor $\tilde K(t)$. The product of the two determines the interference part of the conductance. The ZM factor is suppressed by dephasing, as discussed in Sec.~\ref{sec:deph}.}
\label{fig4}
\end{figure}

Averaging the exponential factor in Eq.~(\ref{KZM1}) over the grand canonical ensemble in the closed ring,
$K_{\rm ZM}(t)$ is obtained
as a sharply peaked periodic function (Fig.~\ref{fig4}).
The height of the peaks equals 1 and their characteristic width $\delta t_{\rm ZM}$ is given by $\delta t_{\rm ZM}=1/\alpha (T\Delta)^{1/2}$.
Notice that the peaks in $K_{\rm ZM}(t)$
appear at precisely the same times as the peaks in the envelope function $\tilde{K}(t)$ and
are much broader, for $T\gg\Delta$, than the latter,
namely $\delta t_{\rm ZM}/\tau_{\rm d}\sim (T/\Delta)^{1/2}$. As a consequence, the effect of the ZM dynamics in the spinful case is masked by SCS and can be neglected in the closed ring. This is in contrast to the spinless case \cite{dmitriev10}, where the counterpart of $K_{\rm ZM}(t)$ is responsible for important changes in the spectrum of the ring, ultimately giving rise to peculiar interaction-controlled AB oscillations \cite{dmitriev10}.

To summarize the results of this section, we have found that the product of the SCS and ZM factors in Eq.~\eqref{25} is
a periodic function with the period $\pi/\alpha\Delta$ (half of that for spinless electrons) imposed by $\tilde K(t)$, see Fig.~\ref{fig4}.
Importantly, this function (which determines the interference part of the conductance)
does not decay with time in a closed ring.

\subsection{SCS dynamics: Tunnel-coupled ring}
\label{s3b}

Now we take into account a finite tunneling coupling of the ring to the leads.
In the noninteracting case, in the close vicinity of $\phi=1/2$ and for $\Gamma_0\ll\Delta$
one can neglect backscattering at the contacts (see Appendix~\ref{App:Dyson}) and then simply replace
$G_R^\pm(\epsilon,L/2)$ with $G_{R0}^\pm(\epsilon+i\Gamma_0/2,L/2)$ \cite{shmakov13},
where $G_{R0}^\pm(\epsilon,L/2)$ is the Green function in a closed ring.
This replacement introduces a factor $\exp(-\Gamma_0t)$ in the integrand of Eq.~\eqref{19a}.
As we demonstrate below, the tunneling coupling in the SCS case is essentially different in
that it cannot be characterized by a single tunneling rate, the same for each energy level.

In order to introduce tunneling into the SCS dynamics, it is convenient to return to the $(\epsilon,x)$ representation,
as in the noninteracting case.
For $|\delta\phi|\ll 1$ and $\Gamma_0\ll\Delta$, we neglect backscattering at the contacts and represent
$G_R^{\pm}(\epsilon,L/2)$ in the presence of SCS as follows (see Appendices \ref{App:Dyson} and \ref{properties}):
\be
G_R^{\pm}(\epsilon,L/2)\simeq\frac{G_{R0}^{\pm}(\epsilon,L/2)}{1+i(\Gamma_0L/2)G_{R0}^{\pm}(\epsilon,L)}~,
\label{Green-eps-Gamma}
\ee
where
\be
G_{R0}^{\pm}(\epsilon,L/2)=\int\!dt\,{\cal G}_{R}^{\pm}(t,L/2)e^{i\epsilon t}~,
\label{Green-eps}
\ee
${\cal G}_{R}^{\pm}(t,L/2)$ is given by Eq.~(\ref{Green-finite}), and a similar expression holds for $G_{R0}^{\pm}(\epsilon,L)$.
One way of thinking about the meaning of Eq.~\eqref{Green-eps-Gamma} is in terms of its expansion in powers of $\Gamma_0$,
which is a sum over winding numbers of the paths that connect the opposite leads and are ``damped" by the possibility of tunneling out of the ring.

In the limit of weak tunneling, the main contribution to the conductance in Eq.~(\ref{1}) comes from $\epsilon$ that are close to the energy levels of the isolated ring. Therefore, we can replace $G_R^{\pm}$ with an auxiliary Green function $\bar G_R^{\pm}$ whose $\epsilon$ dependence coincides with that of $G_R^{\pm}$ in the vicinity of the poles of $G_R^{\pm}$. It is notable that the $t$ dependence of ${\cal G}_{R}^{\pm}(t,L/2)$ is similar to that of $K(t)$ in Sec.~\ref{s3a}, namely both functions are peaked around the times $\pi k/\alpha\Delta$ [Eq.~(\ref{34})]. It follows that, in contrast to the noninteracting ring, where the poles of the Green function at $\epsilon=n\Delta$ are characterized by a single quantum number $n$, the poles in the interacting ring are enumerated by a set of two indices, $n$ and $m$ in the notation of Sec.~\ref{s3a}. Namely the poles occur at $\epsilon=\varepsilon_{nm}$, where
\be
\varepsilon_{nm}\simeq n\Delta +2\alpha m\Delta~.
\label{fine-structure}
\ee
To derive this equation, we notice that the spin and charge factors in $g(t)$ [Eq.~\eqref{g1}] are periodic functions of time with the periods $2\pi/\Delta$ and  $2\pi/\Delta_u$, respectively. Therefore, the energy levels of the system can be written as $n_1\Delta +n_2\Delta_u$ with integer $n_1$ and $n_2$ (see Appendix \ref{properties}). Using Eq.~\eqref{uv} for $\alpha\ll 1$, we obtain Eq.~(\ref{fine-structure}) with
\be
n=n_1+n_2~,\quad m=n_2~.
\label{nm0}
\ee
That is, the levels acquire a fine structure because of SCS, with the ``sublevels" enumerated by $m$.

As shown in Appendix~\ref{properties}, the weight of a sublevel in the Green function vanishes with increasing $|m|$ for $|m|\gg T/\Delta$. Effectively, for the purpose of visualizing resonant transport through the ring for $T\gg\Delta$, one can think of each level of the noninteracting system being split into $T/\Delta$ sublevels. More specifically, the calculation presented in Appendix \ref{properties} yields
\be
\hspace{-1mm}\bar G_R^{\pm} (\epsilon, L/2) =
\frac{\sqrt{\Delta\Delta_u}}{T L}\sum\limits_{nm}
\frac{(-1)^N \lambda_{nm} }{\epsilon -  \varepsilon_{n m} -  \delta E^{\pm}_{\rm ZM} +i  \Gamma_{nm}/2 }~,
\label{GR-pole}
\ee
where $N=N_\pm+1+n$, the structural factor
\be
\lambda_{nm}=\Lambda\left(\frac{n_1\Delta }{T},\,\frac{n_2\Delta_u}{T}\right)
\label{lambda-n1n2}
\ee
with $n_{1,2}$ from Eq.~(\ref{nm0}) is expressible in terms of the dimensionless function
\BEA
\Lambda (x_1, x_2)&=&\left |\Gamma\left(\frac34 +\frac{i x_1 }{2\pi }\right)\right|^{-2}
\left
|\Gamma\left(\frac34 +\frac{ix_2}{2\pi}\right)\right|^{-2}
\nonumber\\
&\times&\frac{\displaystyle\cosh\left(\frac{x_1+x_2}{2}\right)}{\displaystyle \cosh\left(x_1\right)\,\cosh\left(x_2\right)}~,
\label{lambda-nm}
\EEA
and the sublevel broadening $\Gamma_{nm}$ is related to $\lambda_{nm}$ by
\begin{equation}
\Gamma_{nm}=\lambda_{nm}\,\Gamma_0\,\frac{\sqrt{\Delta\Delta_u}}{T}~.
\label{Gamma-nm}
\end{equation}
The function (\ref{lambda-nm}) obeys
\be
\int\!dy\,\Lambda (y,x-y)=1~.
\label{prop}
\ee

The physics of SCS manifests itself through the splitting of the energy levels, with the sublevel spacing $2\alpha\Delta$ in Eq.~(\ref{fine-structure}), and through the $m$-dependence of $\lambda_{nm}$ and $\Gamma_{nm}$. The dependence of these quantities on $n$ does not affect the results for the conductance qualitatively (see Sec.~\ref{s4c1}). Moreover, as we will demonstrate below [Eq.~\eqref{sum}], the classical term in the conductance contains the sum $\sum_m\lambda_{nm}$ and thus---because of the property \eqref{prop}---is not sensitive to the dependence of $\lambda_{mn}$ on $n$ even quantitatively. Therefore, for the sake of simplicity, in the discussion in the rest of this section and in the beginning of Sec.~\ref{s4}, we set $n=0$:
$$\lambda_{nm}\to\lambda_m\equiv\lambda_{0m}~,\quad \Gamma_{nm}\to\Gamma_m\equiv\Gamma_{0m}~.$$ Neglecting the dependence of $\lambda_{nm}$ on $n$ allows us to find a simple generalization of Eq.~\eqref{K18b}, which represents $K(t)$ as a sum of the delta functions, for the case of a slightly open ring, with tunneling fully encoded in the dynamics of the envelope $\tilde K(t)$ [Eq.~(\ref{KK0})].

Within this ``$m$-approximation,'' we obtain for $\alpha\ll 1$:
\be
\bar G_R^{\pm} (\epsilon, L/2)\sim
\frac{{\Delta }}{ T L}\sum\limits_{nm}
\frac{(-1)^n~\lambda_m}{\epsilon -n\Delta -2\alpha m\Delta\mp\Delta\phi +i\Gamma_m/2}~,
\label{GR-pole-m}
\ee
where
\begin{eqnarray}
\Gamma_m&=&\Gamma \lambda_m
\label{gamma-m}
\end{eqnarray}
with
\begin{eqnarray}
\Gamma&=&\Gamma_0 \frac{\Delta }{T}~.
\label{Gamma}
\end{eqnarray}
The weight $\lambda_m$ is of the order of unity for $m=0$ and decays on the characteristic scale of $|m|$ given by
\be
\delta m=T/\Delta~.
\label{69}
\ee
In Eq.~(\ref{GR-pole-m}), we kept only the flux-dependent term in $\delta E^{\pm}_{\rm ZM}$. As seen from Eqs.~\eqref{gamma-m} and \eqref{Gamma}, the tunneling  width of the sublevels is suppressed compared to $\Gamma_0$ by a factor of the order of $\Delta/T$.

To establish a link between Eq.~(\ref{GR-pole-m}) and the calculation in the time domain (Sec.~\ref{s3a}), let us write down the Fourier transform ${\bar{\cal G}}_R^{\pm} (L/2,t)$ of $\bar G_R^{\pm} (\epsilon, L/2)$ Within the $m$-approximation, we have
\be
{\bar{\cal G}}_R^{\pm} (L/2,t)\sim v^{-1} g(t) e^{\pm i \Delta \phi t},
\label{GR-pole-m-t}
\ee
where $g(t)$ [Eq.~(\ref{22})] comes from the plasmon-spinon part of the Green function and is given by
\be
\label{gm}
g(t)\!\sim\! -\frac{i \Delta}{T} \sum \limits_{n} \!\delta\! \left[  t\!-\! \frac{2\pi}{\Delta}\! \left(\!n\!+\!\frac12 \! \right)\right]
 \! \sum\limits_m
\lambda_m e^{-(2i\alpha m\Delta  +\Gamma_m/2)t}~.
\ee
As already mentioned above, the $m$-approximation is particularly convenient to introduce the tunneling-induced decay into the dynamics of SCS. Specifically, this means that, substituting Eq.~\eqref{gm} in Eq.~\eqref{12}, we find $K(t)$ in the $m$-approximation to be exactly given by a sequence of delta functions [as in Eq.~\eqref{K18b}]. The envelope function of the sequence is
\be
\tilde K(t) \sim \left(\frac{ \Delta}{T}\right)^2 \left|\sum\limits_{m} \lambda_me^{-2i\alpha\Delta m t} e^{-\Gamma_m t/2 }\right|^2.
\label{env}
\ee
Apart from showing oscillations in time with a period $\pi/\alpha\Delta$, similar to those in Eq.~(\ref{KK0}), the function $\tilde K(t)$ in Eq.~(\ref{env}) decays---because of tunneling. As was already noted at the beginning of Sec.~\ref{s3b}, the decay of $\tilde{K}(t)$ in the presence of SCS is not characterized by a single tunneling rate: the partial decay rate in channel $m$ in Eq.~(\ref{env}) is given by $\Gamma_m$ [Eq.~(\ref{gamma-m})].

Within the $m$-approximation, the ZM factor [Eq.~(\ref{KZM})]
\be
K_{\rm ZM}(t)\to 1
\label{KZM-m}
\ee
at the times $t=2\pi (n+1/2)/\Delta$ at which the function $g(t)$ is peaked. As a result, the ZM part in Eq.~(\ref{19a}) at $\tau=0$, Eq.~(\ref{FZM}), takes the form
\be
\label{FZM-m}
F_{\rm ZM}(t,0,\phi) \simeq 2\,[\,1+\cos(2\phi\Delta t)\,]~.
\ee

While the $m$-approximation produces the peaks in $g(t)$ [Eq.~\eqref{gm}] and, consequently, in $K(t)$ ``directly" in the form of the exact delta functions, a more accurate calculation (see Appendices \ref{properties} and \ref{App:C}) yields the peaks of nonzero width, having the double-horn structure, as shown in the inset in Fig.~\ref{fig3}.

\section{Calculation of the conductance}
\label{s4}

In this section, we evaluate the AB conductance through the ring in the presence of SCS. We start with the $m$-approximation (Sec.~\ref{s3b}), which, again, fully accounts for SCS and captures all the parametric dependencies. The main reason why we use this approximation for the purpose of illustration is that---as discussed in Sec.~\ref{s3b}---the tunneling dynamics is then fully incorporated in the envelope function $\tilde K(t)$ (the substantially more involved calculation of the classical and quantum contributions to the conductance with the numerical coefficients beyond the $m$-approximation is relegated to Appendix~\ref{beyond-m}).

Substituting $K_{\rm ZM}$ from Eq.~(\ref{KZM-m}) and $K(t)$ in the form of Eq.~(\ref{K18b}) with the envelope function $\tilde K(t)$ from Eq.~(\ref{env}) in Eq.~(\ref{25}), we obtain the transmission coefficient ${\cal T}(\phi)$ as
\begin{align}
\nonumber  {\cal T}(\phi)
&\!\sim\!\frac {\Gamma^2}{\Delta} {\rm Re} \!\!\sum \limits_{mm'}\!  \lambda_m\lambda_{m'}\!
\left[\frac{1}{ (\Gamma_m\!+\!\Gamma_{m'})/2 +2i\alpha \Delta (m-m')}  \right.
\nonumber
\\
&-\!\left.\frac{1}{ (\Gamma_m\!+\!\Gamma_{m'})/2+2i\alpha \Delta (m-m') +2i\Delta\delta\phi}\right],
\nonumber\\
\label{inter}
\end{align}
where $$\delta\phi=\phi-1/2~.$$
It is worth noting that the minus sign in front of the interference term is due to $1/2$ in the argument of the delta function in \eqref{K18b}, which should, therefore, be retained throughout the calculation even for large $n$. Note also that the interference part of ${\cal T}(\phi)$ in Eq.~(\ref{inter}) does not contain any dephasing. We will discuss dephasing in Sec.~\ref{s4c} at the phenomenological level and calculate the dephasing rate, for the case of generic interactions, in Sec.~\ref{sec:deph}.

Equation~\eqref{inter} can also be derived in a different way, by directly substituting the Green function obtained in the $m$-approximation in the $(\epsilon,x)$ representation, Eq.~\eqref{GR-pole-m}, in Eq.~\eqref{15}. We use this method of calculation in Appendix \ref{beyond-m}, where we go beyond the $m$-approximation and find the explicit analytical expression for $\lambda_{nm}$. The transmission coefficient is then obtained by a substitution of Eq.~\eqref{GR-pole} in Eq.~\eqref{15}.

\subsection{Classical term}
\label{s4a}

Assuming a weak tunneling coupling to the leads,
$\Gamma_0 \ll \alpha T$,
we have $\Gamma \ll \alpha \Delta$ and hence can neglect the terms with  $m \neq m'$ in the classical term (the first term in the square brackets) in Eq.~(\ref{inter}). Setting $m'=m$ and replacing the sum over $m$ with an integral, we get a large factor $\int\!dm\,\lambda_m\sim T/\Delta$. As a result, we obtain ${\cal T}_c$ to be of the same order as the transmission coefficient for a noninteracting system:
$
{\cal T}_{c} \sim \Gamma_0/\Delta~.
$

This simply stated result is actually rather nontrivial in view of the SCS-induced energy level splitting in Eq.~(\ref{fine-structure}). For a given noninteracting level, the characteristic number of sublevels participating in resonant transport is given by $\delta m \sim T/\Delta$ [Eq.~(\ref{69})]. However, since each sublevel has the tunneling-induced width of the order of $\Gamma=\Gamma_0 \Delta/T$ [Eq.~(\ref{Gamma})], the total transmission coefficient, which is proportional to the number of channels and the tunneling rate per channel, ${\cal T}_c \sim \delta m\times \Gamma/\Delta \sim \Gamma_0/\Delta$, is the same by order of magnitude. In fact, it does not change at all, because of the ``sum rule" (\ref{prop}), as shown by the calculation of ${\cal T}_c$ beyond the $m$-approximation in Appendix~\ref{beyond-m}. The resulting expression is identical to the classical transmission coefficient for the spinless case \cite{dmitriev10}:
\be
{\cal T}_c=\frac{\pi \Gamma_0 }{\Delta }~.
\label{Drude}
\ee

The insensitivity of the classical conductance to SCS in Eq.~(\ref{Drude}) deserves a special comment and can be understood as follows. The \textit{dc} conductance is proportional to the product of the tunneling rate $\Gamma_0$  for entering the ring  and the probability $W$ to exit the ring (namely the integral over $t$ of the probability density per unit time to escape into the leads). The former ``does not know''  about SCS, since tunneling into the ring weakly tunnel-coupled to the leads happens almost instantly, while the separation of spin and charge takes much longer time $\tau_\text{d}$. On the other hand, $W=1$, since sooner or later the electron should exit the ring (although SCS strongly modifies the escape probability per unit time, it cannot change the total integral over time).  The conservation of the total escape probability translates into the sum rule in the energy representation  in the following way (see Appendix~\ref{beyond-m} for details). The structural factors $\lambda_{nm}$ enter linearly {\it both} the numerator of the Green function [Eq.~(\ref{GR-pole})] and the width $\Gamma_{nm}$ of the quantum levels split by SCS [Eq.~(\ref{Gamma-nm})].  As a result, the partial contribution to the classical conductance  of the $n$-th level  is proportional to $\sum_m\lambda_{nm}^2/\Gamma_{nm}\propto\sum_m\lambda_{nm}$ and, therefore, remains unchanged \cite{litinski16} because of the sum rule, Eq.~\eqref{prop}.

\subsection{Interference term within the $m$-approximation}
\label{s4c}

We now turn to the interference part of the transmission coefficient, given by the second term in the square brackets in Eq.~(\ref{inter}). As was already mentioned at the beginning of Sec.~\ref{s2} and will be demonstrated in Sec.~\ref{sec:deph} below, ZM dephasing is absent in the isotropic model, only emerging because of anisotropy of interactions. In this section, we introduce a phenomenological (external) dephasing rate $\Gamma_\varphi$, in order to illustrate, with a simple example, the interplay between SCS and dephasing. Moreover, we assume that $\Gamma_\varphi\gg\Gamma$. In this case, the calculation simplifies in that the relaxation of the product $K(t)K_{\rm ZM}(t)$ is characterized by a single relaxation rate, in contrast to Eq.~(\ref{env}). We can, therefore, describe the SCS dynamics by $\tilde K(t)$ in a closed ring [Eq.~(\ref{env}) with $\Gamma_m=0$] and associate the factor $\exp (-\Gamma_\varphi t)$ with $K_{\rm ZM}(t)$ (as in Fig.~\ref{fig4}). As will be seen in Sec.~\ref{sec:deph}, the condition $\Gamma_\varphi\gg\Gamma$ is indeed generically satisfied for ``internal" ZM dephasing, characterized, then, by the simple exponential decay.

For $\Gamma_m=0$, replacing the summation over $m$ in Eq.~(\ref{env}) with an integral yields $K_0(t)$ in the form
\be
K_0(t) \sim \left(\frac{ \Delta}{T}\right)^2 \left|\int dm \, \lambda_m \, e^{-2i\alpha\Delta m t} \right|^2~.
\label{envK0}
\ee
Using $K_0(t)$ from Eq.~(\ref{envK0}) in Eq.~(\ref{K18}) and substituting the resulting $K(t)$ in Eq.~(\ref{25}), together with $K_{\rm ZM}(t)=\exp (-\Gamma_\varphi t)$, we obtain
\be
{\cal T}_q \sim - \frac{\Gamma_0^2}{\Delta}~ {\rm  Re} \sum \limits_k \int\limits_0^\infty  dt  K_0(t-t_k) e^{-\Gamma_\varphi t}e^{-2i\Delta\delta\phi t}~,
\label{77}
\ee
where $t$ in the integral is understood as a ``coarse-grained" time $2\pi n/\Delta\to t$. Note that $\Gamma_\varphi$ stands in Eq.~(\ref{77}) in a combination  $\Gamma_\varphi+2i\Delta\delta\phi$, which translates into the addition of $\Gamma_\varphi$ in the denominator of the interference term in Eq.~(\ref{inter}). For $\Gamma_\varphi \ll \alpha T$, we obtain
\be
{\cal T}_q \sim -\frac{\Gamma_0^2}{\alpha T\Delta }~{\rm Re}~ \frac{\kappa(\delta\phi)}{1- \exp\left[ \frac{\pi}{\alpha \Delta   }
\left(-\Gamma_\varphi -2i\Delta \delta\phi\right) \right]}~,
\label{T-int-0}
\ee
where
\be
\kappa(\delta \phi)\!=\!\alpha T\!\!\! \int dt  e^{-2i\Delta \delta \phi t} K_0(t) \sim \frac{\Delta}{T}\!\! \int\!\! dm\, \lambda_m\lambda_{m-\delta\phi/\alpha}
\label{k}
\ee
is of the order of unity at $\delta\phi=0$ and decays on the scale $\delta\phi \sim \alpha T/\Delta$. We see that the function ${\cal T}_q(\delta \phi) $ shows series of narrow negative resonances of width $\Gamma_\varphi/\Delta$. The distance between resonances is given by $ \alpha$, while their amplitudes are modulated by the slowly decaying function $\kappa(\delta\phi)$. A more accurate calculation of ${\cal T}_q$, performed beyond the $m$-approximation, is presented in Appendix \ref{beyond-m} and yields Eq.~\eqref{T-int-1} which does not differ from Eq.~(\ref{T-int-0}) qualitatively.

\subsection{AB conductance for the isotropic model}
\label{s4c1}

Combining Eqs.~\eqref{Drude} and \eqref{T-int-1}, we arrive at the expression for the tunneling conductance in the form \cite{asymmetric}
\be
{\rm G}(\phi)\!=\!\frac{e^2\Gamma_0}{\Delta}\left[1- \frac{\Gamma_0}{2T} \sum\limits_k b_k \frac{\Gamma_\varphi/2\Delta}{(\delta\phi-\alpha k)^2+ (\Gamma_\varphi/2\Delta)^2 } \right],
\label{finalG}
\ee
where
\be
b_k=\frac{1}{4\cosh^2(\Delta k/2T)}~.
\ee
Comparing Eq.~\eqref{finalG} with the corresponding expression for the spinless case  (see footnote [24] in Ref.~\cite{dmitriev10}), we see that the interference part of the conductance is given in both cases by the sum of the thermodynamically weighted Lorentz peaks: with the ZM Gibbs factor in the spinless case and with $ (\p f/ \p \epsilon)_{\epsilon=\Delta k}$ in the spinful case. It is worth stressing that the exact shape of the envelope of the resonances (the dependence of the weights $b_k$ on $k$) fully accounts for the fine (double-horn) structure of the peaks in $K(t)$, which was discarded within the $m$-approximation.

\begin{figure}[ht!]
\leavevmode \epsfxsize=8.5cm
\centering{\epsfbox{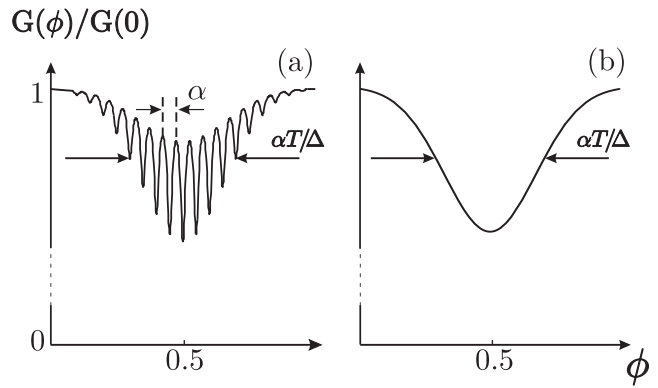}}
\caption{ Tunneling conductance of the ring for (a) $\Gamma_\varphi\ll\alpha\Delta$ and (b) $\Gamma_\varphi\gg\alpha\Delta$.}
\label{fig5}
\end{figure}

The function ${\rm G(\phi)}/{\rm G(0)}$ is schematically plotted in Fig.~\ref{fig5}. For $\Gamma_\varphi \ll \alpha \Delta$, the conductance as a function of $\delta \phi$ shows a series of narrow peaks of width $\Gamma_\varphi/\Delta$ separated by a distance $\alpha$ (Fig.~\ref{fig5}a). The splitting of the resonance in ${\rm G(\phi)}$ is a direct consequence of SCS. As the dephasing rate $\Gamma_\varphi$ increases and becomes larger than $ \alpha \Delta$, the resonances in ${\rm G}(\phi)$ overlap and form a single peak of width $\alpha T/\Delta$ (Fig.~\ref{fig5}b). It is worth noting that if one customarily extracted the dephasing rate from the function ${\rm G}(\phi)$ as the width of this single peak, one would arrive at the erroneous conclusion that the dephasing rate is given by $\alpha T$, the inverse single-particle lifetime in a homogeneous spin-charge-separated Luttinger liquid (cf.\ a similar situation in the spinless case in Ref.~\cite{dmitriev10}).

It is interesting to notice that the noninteracting conductance is obtained from Eq.~\eqref{finalG} by sending $\alpha \to 0$, despite the fact that the condition of the derivation of Eq.~(\ref{finalG}) is $\alpha\gg\Gamma/\Delta$ (see the discussion in the beginning of Sec.~\ref{s4a} and in Appendix~\ref{beyond-m}). Indeed, neglecting $\alpha k $ in the denominator of Eq.~\eqref{finalG}, replacing the sum over $k$ with an integral, and substituting $\Gamma_\varphi$ with $\Gamma_0$, we reproduce the noninteracting transmission coefficient, Eq.~\eqref{small-gamma}. This is because, for $\alpha \Delta\ll \Gamma$ (or, equivalently, $\alpha T \ll \Gamma_0$), the SCS-induced level splitting is smaller than the level broadening and the suppression of $K(t)$ by tunneling occurs already within the first bunch of peaks, implying that SCS has no time to develop ($\Gamma_0\tau_{\rm d}\gg 1$).

From Eq.~\eqref{finalG} we conclude that at $\phi=1/2$ the ratio of the quantum (${\rm G}_q$) and classical (${\rm G}_c$) contributions to the conductance
is given by
\be
\frac{{\rm G}_q}{{\rm G}_c}=\frac{{\cal T}_q}{{\cal T}_c} = - \frac{\Gamma}{\Gamma_\varphi} \ll1~, \qquad \delta\phi=0~.
\label{-1}
\ee
It is instructive to check what happens with the conductance if the external dephasing rate exactly equals zero. In this case, the tunneling rates $\Gamma_{nm}$ should be retained in the denominator of Eq.~\eqref{58}, which makes it impossible to represent ${\cal T}_q$ as a single sum over $k$. One can, however, still calculate the conductance at $\delta\phi=0$, along the lines of the calculation in Appendix \ref{beyond-m}, with the result
\be
\frac{{\rm G}_q}{{\rm G}_c}=\frac{{\cal T}_q}{{\cal T}_c} \to - 1~, \quad\quad \delta\phi=0~,
\label{-1-1}
\ee
which implies almost exact destructive interference at $\delta\phi=0$. This exemplifies the notion that SCS {\it by itself} does not suppress quantum interference. Specifically, SCS does not lead to any decay of $K(t)$ [see, e.g., Eqs.~(\ref{env}) and (\ref{K-exact}), in which the decay is only due to tunneling] and thus does not produce any dephasing. Moreover, in contrast to a naive expectation, the SCS-induced splitting does not lead to any suppression of the interference pattern, either [Eq.~(\ref{-1-1})].

\subsection{Generic model}
\label{sec:deph}

Let us now return to the discussion of a generic (anisotropic in spin and chirality spaces) model characterized by four distinct interaction constants $\alpha_{2\|},\alpha_{2\perp},\alpha_{4\|}$, and $\alpha_{4\perp}$.  As we mentioned above, the violation of isotropy in spin-chirality space leads to two important effects, both coming from ZM fluctuations. First, {\it thermodynamic} fluctuations of the ZM  population numbers lead to an additional splitting of the AB resonances compared to Eq.~\eqref{finalG} (``inhomogeneous broadening"). Second, {\it dynamical} fluctuations of these numbers, arising from the tunneling exchange of electrons with the leads, give rise to dephasing (``homogeneous broadening"), known as ZM dephasing \cite{dmitriev10}.

\subsubsection{ZM splitting}

In the generic case, the ZM energy reads \cite{pletyukhov06,meden08}:
\begin{align}
&E_{\rm ZM}=\frac{\Delta_{\rm F}}{8 v_{\rm F}}
\label{Ezm1}
\\
\nonumber
&\times \left [ v_{\rm c}(J_c - 4\phi)^2+v_{\rm s} J_s^2 + u_{\rm c} (N_{\rm c} -4N_0)^2 +u_{\rm s} N_{\rm s}^2 \right]~,
\end{align}
where $N_c$ and $J_c$ are given by Eqs.~(\ref{N_c}) and (\ref{J_c}),
\BEA
N_{\rm s}&=&N_+^\uparrow-N_+^\downarrow +(N_-^\uparrow-N_-^\downarrow)~,\\
J_{\rm s}&=&N_+^\uparrow-N_+^\downarrow -(N_-^\uparrow-N_-^\downarrow)~,
\EEA
and
\BEA
u_{\rm c}&=& v_{\rm F} \left[ 1+(\alpha_{4\|}+\alpha_{4\perp})+(\alpha_{2\|}+\alpha_{2\perp})\right]~,
\\
u_{\rm s}&=& v_{\rm F} \left[ 1+(\alpha_{4\|}-\alpha_{4\perp})+(\alpha_{2\|}-\alpha_{2\perp})\right]~,
\\
v_{\rm c}&=& v_{\rm F} \left[ 1+(\alpha_{4\|}+\alpha_{4\perp})-(\alpha_{2\|}+\alpha_{2\perp})\right]~,
\\
v_{\rm s}&=& v_{\rm F} \left[ 1+(\alpha_{4\|}-\alpha_{4\perp})-(\alpha_{2\|}-\alpha_{2\perp})\right]~.
\EEA
The level spacing $\Delta_F=2\pi v_F/L$ is determined by the bare (noninteracting) Fermi velocity $v_F$. The charge and spin plasmonic excitations propagate with the velocities $u=\sqrt{u_{\rm c}v_{\rm c}}$ and $v=\sqrt{u_{\rm s}v_{\rm s}}.$  Neglecting the terms in $u$ and $v$ that are quadratic in the coupling constants, we find
\BEA
u&\simeq& v_{\rm F}(1+\alpha_{4\|}+\alpha_{4\perp})~,
\label{u87}
\\
v&\simeq& v_{\rm F}(1+\alpha_{4\|}-\alpha_{4\perp})~.
\label{v88}
\EEA
Note that the spinon velocity $v$ is, generically, not equal to $v_F$.

The calculation of the conductance for anisotropic interactions proceeds along the lines of the one presented for the isotropic model in Appendix \ref{beyond-m}. The interference part of the conductance ${\cal T}_q$ is given by an expression analogous to Eq.~\eqref{58a}. The spectrum of two-particle excitations that enters the denominator of Eq.~\eqref{58a} is given by the difference of the full energies for right and left movers and can be written as the sum of the ZM and spin-charge parts:
\be
\Delta E=\delta E_{\rm ZM}^{+\uparrow}-\delta E_{\rm ZM}^{-\uparrow}+\varepsilon(n_1,n_2)-\varepsilon(n_1',n_2')~.
\ee
The ZM contribution [cf.\ Eq.~(\ref{C*})] is now given by
\begin{align}
&\frac{\delta E_{\rm ZM}^{+\uparrow}-\delta E_{\rm ZM}^{-\uparrow}}{\Delta_{\rm F}}=J^\uparrow (1+\alpha_{4\|}-\alpha_{2\|})+J^\downarrow (\alpha_{4\perp}-\alpha_{2\perp})
\nonumber
\\
&\hspace{2cm}-2\phi (1+\alpha_{4\|}+\alpha_{4\perp}-\alpha_{2\|}-\alpha_{2\perp})
\label{zm90}
\end{align}
with
\BEA
J^\uparrow &=&N_+^\uparrow-N_-^\uparrow~,
\\
J^\downarrow &=&N_+^\downarrow-N_-^\downarrow
\EEA
being the circular currents for different spin polarizations. The spin-charge part
\begin{align}
&\varepsilon(n_1,n_2)-\varepsilon(n_1',n_2')=\Delta (n_1-n_1')+\Delta_u (n_2-n_2')
\nonumber
\\
\nonumber
&\simeq\Delta_{\rm F}[(1+\alpha_{4\|}-\alpha_{4\perp})(n-n')+2\alpha_{4\perp}(m-m')]
\end{align}
includes the SCS splitting $2\alpha_{4\perp}(m-m')$. Again, similar to the isotropic case, we find that the main contribution to the interference term comes from
$$n'=n+J^\uparrow-1~.$$ Keeping this term only, we find that $\Delta E$ now becomes
\BEA
\Delta E\!&\simeq&\!\Delta_{\rm F}\left[(\alpha_{4\perp}-\alpha_{2\|})J^\uparrow +(\alpha_{4\perp}-\alpha_{2\perp})J^\downarrow\right.\nonumber\\
&-&\!\left.2\delta\phi+2\alpha_{4\perp}(m-m')\right]~.
\label{ZMsplitting}
\EEA
Here we neglected the interaction terms (of the order of $\alpha$) which do not depend on $J^{\uparrow,\downarrow}$ and $m,m'$. It is worth commenting on the structure of Eq.~(\ref{ZMsplitting}) as far as the contributions of different spins are concerned. Recall that, by construction, Eq.~(\ref{ZMsplitting}) is for the energy of two-particle excitations with spin up. The contribution to this energy of the interaction with electrons of the same spin (the term proportional to $J^\uparrow$) contains, however, the coupling constant $\alpha_{4\perp}$, in contrast to the factor in front of $J^\uparrow$ in Eq.~(\ref{zm90}). This is because the spectrum of spinon excitations is renormalized by interactions between electrons with opposite spins [Eq.~(\ref{v88})], hence the substitution of $\alpha_{4\perp}$ for $\alpha_{4\|}$ in the combination $\alpha_{4\|}-\alpha_{2\|}$ coming from ZM.

Two cases are special: the fully isotropic case,
\begin{equation}
\alpha_{2\|}=\alpha_{4\perp}=\alpha_{2\perp}~,
\label{case1}
\end{equation}
 discussed for most of the paper, and the case
\begin{equation}
3\alpha_{4\perp}=\alpha_{2\perp}=\alpha_{2\|}~,
\label{case2}
\end{equation}
in both of which the current-dependent  ZM energy splitting is exactly equal to zero, so that
\begin{equation}
\Delta E\simeq\Delta_F[2 \alpha_{4\perp} (m-m')-2\delta\phi]~.
\label{NoZM}
\end{equation}
That is, in both cases, ZM fluctuations are irrelevant and the only source of splitting of the resonance in ${\rm G}(\phi)$ is SCS.

In the generic case, Eq.~\eqref{ZMsplitting} should be substituted in Eq.~\eqref{finalG} and the result should be averaged with the ZM Gibbs factor. This yields additional splitting of the AB resonances. In the spinless case ($\alpha_{4\perp}=\alpha_{2\perp}=0$), one of the consequences of the ZM-induced splitting is ``persistent-current blockade" \cite{dmitriev10}. Moreover, there exists an inherent link between the ZM splitting and ZM dephasing, as was emphasized in Ref.~\cite{dmitriev10}. A similar mechanism of dephasing induced by ZM fluctuations was also discussed (under the name of ``topological dephasing") in the context of interference of fractional quantum Hall modes in Ref.~\cite{park15}. Importantly, in addition to the ZM splitting, the generic spinful model also gives rise to dephasing caused by a tunneling exchange of electrons between the ring and the leads. Let us now calculate the corresponding dephasing rate.

\subsubsection{ZM dephasing}

As was found in Ref.~\cite{dmitriev10} (see also Ref.~\cite{dmitriev14} for a more detailed discussion), in the spinless case, dephasing is dominated by tunneling-induced temporal fluctuations of the circular current $N_{+}-N_{-}$. These are accompanied by fluctuations of an effective flux [the contributions to $\Delta E$ in Eq.~(\ref{ZMsplitting}) that are associated with the circular currents can be interpreted in terms of an interaction-induced correction to the external flux $\phi$], directly translating into random changes of the electron phase difference $\int_0^t \Delta E(t')dt'$. The resulting dephasing rate for the spinless case is given by \cite{dmitriev10}
\be
\Gamma_\varphi^{(0)}=4\Gamma_0\frac{T}{\Delta}~.
\label{Gamma-phi0}
\ee
The dephasing rate $\Gamma_\varphi^{(0)}$ characterizes the temporal decay of the interaction-induced factor
\be
\exp[-S(t)]= \left\langle  \exp(-i\alpha\Delta \int_0^t dt' [N_{+}(t')-N_{-}(t')]) \right\rangle,
\label{Sphi0}
\ee
where the averaging is performed over noise realizations. An important piece of physics behind Eq.~\eqref{Gamma-phi0} is thus telegraph noise of the occupation numbers for individual energy levels. For a given level, the occupation number flips between 0 and 1 because of tunneling in time $\Gamma_0^{-1}$, thus leading to changes of $N_\pm$ by $\pm 1$ on a time scale of $(\Gamma_0 T/\Delta)^{-1}$, where the factor $T/\Delta$ is the characteristic number of levels (``fluctuators") in the temperature window. It is the latter time scale that determines $\Gamma_\varphi^{(0)}$ in Eq.~(\ref{Gamma-phi0}).

Let us now turn to the spinful case. As far as the fluctuations of the circular current are concerned, the physics looks very much like what we had for spinless electrons. A finite dephasing rate is now due to the tunneling dynamics of two currents $J^{\uparrow}$ and $J^{\downarrow}$. Generalizing Eq.~(\ref{Sphi0}) to the spinful case
[by integrating the ZM dependent part of Eq.~(\ref{ZMsplitting}) over time],
we have
\begin{align}
&\label{Sphi-spinful}
\exp[-S(t)]
\\
\nonumber
&=\left\langle e^{i\Delta_{\rm F}\int_0^t dt'[(\alpha_{4\perp}-\alpha_{2\|})J^\uparrow (t')+(\alpha_{4\perp}-\alpha_{2\perp})J^\downarrow (t')]}\right\rangle~,
\end{align}
where we used $\Delta E$ with time-dependent circular currents.

Importantly, in addition to the straightforward change of the number of fluctuating currents from one to two, telegraph noise in the currents $J^{\uparrow}$ and $J^{\downarrow}$ results in a dephasing-induced broadening of {\it sublevels} in the ring. It is worth reiterating, however, that---despite the emergence of the sublevel structure in the spectrum of spinful electrons---the dephasing action for the spinless case is reproduced from Eq.~(\ref{Sphi-spinful}) by simply setting $\alpha_{4\perp}= \alpha_{2\perp}=0$, $\alpha_{2\|}=\alpha$, and $\Delta=\Delta_F$.
It is also worth noting once again that, in the symmetric cases (\ref{case1}) and (\ref{case2}), the two-particle excitation energy $\Delta E$ does not depend on the ZM numbers [Eq.~\eqref{NoZM}]. It follows, then, that ZM dephasing is ineffective \cite{footnote-ZM} in these cases: $S(t)=0$ in Eq.~(\ref{Sphi-spinful}). In particular, in
the isotropic model (\ref{case1}), this happens because the interaction of the interfering electrons with electrons of the same chirality and opposite spin exactly cancels the phase difference arising from the interaction with electrons of the opposite chirality (which was the source of ZM dephasing in the spinless case).

Remarkably, the dephasing rate does not change (up to a factor of two) after the addition of spin. Indeed, quite generally, the strength of ZM dephasing is determined by the product of the number of active  fluctuators and the ``flipping" rate for a single fluctuator. As demonstrated above, each $n$-th level in the noninteracting ring splits, because of SCS, into sublevels characterized by the quantum number $m$ with $|m|\alt {\rm max}(|n|,T/\Delta)$, see Appendix \ref{properties}. Since $n$  is limited to within the ``thermal" interval $ |n|  \lesssim T/\Delta$, the number of active fluctuators is given by $(T/\Delta)^2$.   On the other hand, we found that the  tunneling rate for a sublevel is suppressed, again because of SCS, by a factor of $\Delta/T$, i.e., the characteristic flipping rate is given by $\Gamma$ [Eq.~(\ref{Gamma})]. As a result, the dephasing rate is the same by order of magnitude as in the spinless case: $\Gamma_\varphi \sim \Gamma \left({T}/{\Delta}\right)^2\sim\Gamma_0 T/\Delta $. In fact, it only differs by a factor $2$ compared to Eq.~(\ref{Gamma-phi0}), where 2 is the number of random fields  $J^\uparrow (t)$ and $J^\downarrow (t)$, i.e., the number of independently fluctuating noise baths:
\be
\Gamma_\varphi = 2\Gamma_\varphi^0= 8\Gamma_0 \frac{T}{\Delta}~.
\label{Gamma-phi}
\ee
The derivation of the numerical prefactor in Eq.~(\ref{Gamma-phi}) is essentially identical to that in Eq.~(\ref{Drude}) [both rely on the sum rule (\ref{prop})], which highlights the intrinsic connection between dephasing and dissipation [cf.\ the factor of 2 in the conductance in Eq.~(\ref{1})]. A remarkable feature of Eq.~\eqref{Gamma-phi} is that $\Gamma_\varphi$ does not depend on $\alpha$, in contrast to the single-particle decay rate in a homogeneous Luttinger liquid, which is given by (for the spinful case) $\alpha T$ \cite{gornyi05,lehur05,yashenkin08}.

\subsubsection{AB conductance}

Let us finally discuss the flux dependence of the conductance for the generic model with anisotropic interaction. Qualitatively, the overall picture of the AB resonances in ${\rm G}(\phi)$ looks much the same as in Fig.~\ref{fig5}---the conductance shows narrow peaks of width $\Gamma_\varphi/\Delta$ with a smooth envelope of width $\alpha T/\Delta$. However, quantitatively, the picture is different. First, the distance between neighboring peaks is no longer constant. Namely there are now three different periods: the period $\alpha_{4\perp}$, coming from SCS, and two additional periods $(\alpha_{4\perp}-\alpha_{2\|})/2$ and $(\alpha_{4\perp}-\alpha_{2\perp})/2$ which come from the ZM splitting. Second, the amplitude of the resonances decreases compared to the isotropic case (for the same dephasing rate) by a factor of $ \Delta/T$, which is a product of two factors $\sqrt{\Delta/T}$ associated, respectively, with thermodynamic fluctuations of $J^\uparrow$ and $J^\downarrow$. Assuming that all three periods are of (the same) order $\alpha$, we find that the resonances are well resolved ($\Gamma_\varphi\ll\alpha\Delta $) for $$\Gamma_0\ll \alpha\Delta^2/T~.$$ It follows that at $\phi=1/2$ the ratio of the quantum and classical contributions to the conductance at $\phi=1/2$ is given by
\be
\frac{{\rm G}_q}{{\rm G}_c}=\frac{{\cal T}_q}{{\cal T}_c}\sim -\left(\frac{\Delta}{T}\right)^3~, \quad \delta\phi=0~.
\label{-3}
\ee
This result should be contrasted with the spinless case, where the corresponding ratio is $(\Delta/T)^{3/2}$. Note that the difference stems not only from the mere difference in the numbers of degrees of freedom [each of two spins produces a ZM factor of $(\Delta/T)^{1/2}$ in Eq.~(\ref{-3})] but also from SCS which gives one additional power of $\Delta/T$, reflecting the thermal averaging over the SCS-induced sublevels with $\Gamma_\varphi\agt\Gamma$.

\section{Conclusions}
\label{s5}

We have discussed the influence of spin-charge separation on transport through the Aharonov-Bohm tunneling interferometer in the limit of high temperature $T$
($T$ much larger than the level spacing $\Delta$). We have demonstrated that spin-charge separation leads to splitting of each energy level of an electron in the ring into a series of $T/\Delta$ sublevels. The tunneling width of the sublevels is suppressed by a factor $\Delta/T$ compared to the tunneling rate for a single level in a noninteracting ring. This factor can be interpreted as the probability of the spin and charge parts of a single-particle excitation meeting each other at the contact. Remarkably, the classical part of the conductance coincides with its noninteracting value. This happens because the number of the tunneling channels (number of ``active" sublevels) increases by a factor $T/\Delta$, thus compensating the decrease of the tunneling rate for a single channel.

We have shown that spin-charge separation in the quantum ring tunnel-coupled to the leads does not lead to an enhancement of the Aharonov-Bohm dephasing rate $\Gamma_\varphi$. Generically, $\Gamma_\varphi$ in the tunneling interferometer does not depend on the strength of electron-electron interactions and is given by the total rate at which the ring exchanges particles with the leads for the noninteracting case $\Gamma_0T/\Delta$. The tunneling conductance as a function of the  dimensionless magnetic flux shows a series of sharp negative peaks having the width $\Gamma_\varphi/\Delta$. The distance between the peaks is controlled by the interaction strength $\alpha$. The peaks are grouped in bunches of width $\alpha T/\Delta$ (Fig.~\ref{fig5}a) centered at half-integer fluxes. With increasing tunnel coupling, each of the bunches transforms into a structureless peak (Fig.~\ref{fig5}b).

Before concluding the paper, it is worth remarking on the case of strong interactions, $\alpha\agt 1$, where the commensurabilty of the spin and charge velocities is of crucial importance \cite{jagla93,hallberg04,meden08,rincon09}. Although our analytical approach, which extensively uses the inequality $\alpha\ll 1$, does not directly apply to that case, we expect that the physics of the problem made apparent in the formalism we developed here does not change dramatically as the strength of interaction increases. As a matter of fact, the ``recombination" of the factorized spin and charge parts of the electron after many revolutions they perform around the ring (Fig.~\ref{fig1}) before escaping to the leads can be viewed as a precursor of the ``strong commensurability" showing up for $\alpha\agt 1$, where the commensurability between the spacing of the sublevels split by SCS and the spacing of the ``bare" levels (Fig.~\ref{F-new}) becomes relevant.

\section{Acknowledgements}

The work was supported by the joint grant of the Russian Science Foundation (Grant No. 16-42-01035) and the Deutsche Forschungsgemeinschaft (Grant No. MI 658-9/1).

\appendix

\section{Dyson equation}
\label{App:Dyson}

In this Appendix, we derive the Dyson equation whose solution is the electron Green function in a noninteracting ring weakly coupled to the contacts. The main purpose of this derivation is to present the results for the noninteracting case in a way that can be easily generalized to take SCS into account.

In the absence of interaction, the retarded Green functions in a closed ring are given by
\begin{align}
&G_{R0}^\pm(\epsilon,x)= -\frac{i}{v}\sum \limits_m \theta (mL \pm x) e^{i q_{\pm}(mL \pm x)}
\nonumber
\\
&=\frac{1}{v} \frac{e^{iq_\pm x}}{i(1-e^{iq_\pm L})+i0}~, \qquad \text{for}\,\,\,  0<x<L~,
\label{GGpm}
\end{align}
where
\be
q_\pm=k\mp\frac{2\pi\phi}{L},\qquad k=\frac{\epsilon}{v}
\ee
and $\pm$ denotes electrons moving clockwise ($+$) and counterclockwise ($-$). The index $0$ means that scattering by contacts is absent (closed ring). In the vicinity of its poles, the Green function can be approximated as
\be
G_{R0}^\pm(\epsilon,x) \simeq \frac{1}{L}\sum\limits_n \frac{e^{2\pi i n x/L}}{\epsilon-\varepsilon_n^\pm +i0}~,
\label{pole}
\ee
where
\be
\varepsilon_n^\pm=\Delta(n\pm\phi)~.
\label{levels1}
\ee

\begin{figure}[ht]
\includegraphics
[width=0.15\textwidth]{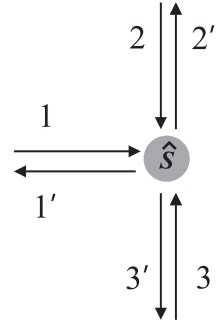}
\caption{Schematic picture of one of two identical contacts: a metallic lead (channels $1$ and $1'$) is connected to the ring with channels ($2,2'$) and ($3,3'$).}
\label{AppFig1}
\end{figure}

We describe the coupling of the ring to two leads by the scattering matrix $\hat{S}$, identical for both contacts, which connects the in-going $(1,2,3)$ and out-going $(1',2',3')$ states (Fig.~\ref{AppFig1}):
\begin{equation}
\hat{S}=\left(
\begin{array}{ccc}
r & t_{\rm out} & t_{\rm out} \\
t_{\rm tun} & t_{\rm b} & t_{\rm in} \\
t _{\rm tun}& t_{\rm in} & t_{\rm b} \\
\end{array}
\right)~.
\end{equation}
For a time-reversal symmetric scattering at the contact, we have $$t_{\rm tun}=t_{\rm out}~.$$ In the limit of weak tunneling ($|t_{\rm tun}|\ll 1 $), however,
$|t_{\rm tun}|=|t_{\rm out}|$ both with and without time-reversal symmetry for scattering at the contact.

For the simplest model, which we consider here, of a point-like contact described by the tunneling Hamiltonian with the tunneling coupling $t_0$, we have \cite{aristov10}
\begin{eqnarray}
\begin{array}{ll}
t_{\rm b}=-\dfrac{\gamma}{1+\gamma}~, & \quad t_{\rm tun}=t_{\rm out}=-i ~{\rm sgn} (t_0)  \dfrac{\sqrt{2\gamma}}{1+\gamma}~,\\
&\\
t_{\rm in}=\dfrac{1}{1+\gamma}~, & \quad r=\dfrac{1-\gamma}{1+\gamma}~,
\end{array}
\label{tunnHam}
\end{eqnarray}
where
\be
\gamma=\frac{2|t_0|^2}{v^2}
\label{a7}
\ee
is the dimensionless tunneling transparency related to the tunneling rate $\Gamma_0$ by
\be
\Gamma_0=\frac{2\gamma\Delta}{\pi}~.
\ee
In Eq.~(\ref{a7}), we assumed that the density of states in the leads equals that in the ring.

The amplitude $t(\epsilon,\phi)$ of transmission through the ring tunnel-coupled to the leads can be calculated as
the product of three transfer matrices, one corresponding to the ring and two others corresponding to two contacts. In this way we obtain, for the case of symmetrically (with the arms of the interferometer of equal length) placed contacts,
\BEA
&&t(\epsilon, \phi)=t_{\rm out} t_{\rm in} \left[\frac{t_{\rm b}-t_{\rm in}+e^{-ik L/2}\cos(\pi\phi)}{t_{\rm in}^2-t_{\rm b}^2
+e^{-ik L}-2t_{\rm in}e^{-ik L/2}\cos(\pi\phi)} \right.
\nonumber\\
&&\left.\qquad - \frac{t_{\rm b}-t_{\rm in}-e^{-i k L/2}\cos(\pi\phi)}{t_{\rm in}^2-t_{\rm b}^2+e^{-i k L}+2t_{\rm in}e^{-i k L/2}\cos(\pi \phi)} \right]~,
\label{t-e-fi-1}
\EEA
For the case of the tunneling Hamiltonian \eqref{tunnHam}, this can be rewritten as
\BEA
\nonumber
t(\epsilon, \phi)&=&-\frac{2\gamma}{(1+\gamma)^2} \left[\frac{Z(\phi)e^{i(kL/2- \pi \phi)} }{1-\tilde t_{\rm in}^2(\phi)e^{i(kL-2\pi\phi)}}\right.
\\
&+& \left.\frac{Z^*(\phi)e^{i(kL/2+ \pi \phi)} }{1-\tilde t_{\rm in}^{*2}(\phi)e^{i(kL+2\pi\phi)}}\right]~,
\label{t-e-fi-2}
\EEA
where
\be
\tilde t_{\rm in}(\phi)=\frac{e^{i\pi \phi} (1-\gamma)}{\cos \pi\phi +i\sqrt{\sin^2\pi\phi+\gamma^2} }~,
\label{tilde-t}
\ee
and
\BEA
Z(\phi)&=&\frac{\gamma \cos \pi\phi +i\sqrt{\sin^2\pi\phi+\gamma^2}}{i\sqrt{\sin^2\pi\phi+\gamma^2}}
\nonumber
\\
&\times& \frac{e^{i\pi\phi} (1+\gamma)}{\cos \pi\phi +i\sqrt{\sin^2\pi\phi+\gamma^2}}~.
\label{Sphi}
\EEA
From Eq.~\eqref{t-e-fi-2}, we have for the transmission coefficient averaged over $\epsilon$ \cite{dmitriev10}:
\be
{\cal T}(\phi)=\langle  |t(\epsilon, \phi )|^2\rangle_\epsilon=\frac{2\gamma \cos^2\pi \phi}{\gamma^2+\cos^2\pi \phi}~.
\ee
In the limit of weak tunneling ($\gamma \ll 1$), ${\cal T}(\phi)$ shows a sharp antiresonance at $\phi=1/2$:
\be
{\cal T}(\phi)\simeq \frac{2\gamma \pi^2 \delta \phi^2}{\gamma^2+\pi^2 \delta \phi^2}~.
\label{small-gamma}
\ee

Having in mind to generalize this approach to the case of SCS, we notice that, for $\gamma \ll 1$, one can neglect backscattering by the contacts (except in the close vicinity of $\phi=0$ with $|\phi|\alt\gamma$) \cite{dmitriev10}. Indeed, from Eqs.~\eqref{tilde-t} and \eqref{Sphi} we find for $\gamma\to 0$ that
\be
\tilde t_{\rm in}(\phi) \to 1-\gamma \simeq t_{\rm in},  \quad Z(\phi) \to 1.
\label{limit}
\ee
In this limit and except in the narrow interval of $\phi$ around $\phi=0$, Eq.~\eqref{t-e-fi-2} yields a sum of the transmission amplitudes for clockwise-  and counterclockwise-moving electrons (with different winding numbers), with backscattering playing no role:
\BEA
\nonumber
t(\epsilon, \phi)&\simeq&-2\gamma \left[\frac{e^{i(kL/2- \pi \phi)} }{1- (1-\gamma)e^{i(kL-2\pi\phi)}}\right.
\\
&+& \left. \frac{e^{i(kL/2+ \pi \phi)} }{1-(1-\gamma)e^{i(kL+2\pi\phi)}}\right]~.
\label{t-e-fi-3}
\EEA
It is worth stressing that the exact formula  \eqref{t-e-fi-2} differs from Eq.~\eqref{t-e-fi-3} only by the renormalization  $t_{\rm in}\to t_{\rm in}(\phi)$ and by the appearance of the factor $Z(\phi)$. Averaging $ |t(\epsilon, \phi)|^2$ over energy with the transmission amplitude given by Eq.~\eqref{t-e-fi-3} and taking
the limit $\delta \phi \ll 1$, we recover Eq.~\eqref{small-gamma}.

It is useful to derive Eq.~\eqref{t-e-fi-3} in a different way, by means of the Dyson equation. Although such a derivation is more complicated, it has the advantage that it can easily be generalized to the interacting case. As demonstrated above, for small $\gamma$, backscattering by the contacts can be neglected, i.e., the Green functions of electrons propagating along the ring in the clockwise and counterclockwise directions can be calculated independently. Let us focus on the calculation of the retarded Green function of clockwise-moving electrons.

\begin{figure}[ht]
\includegraphics
[width=0.4\textwidth]{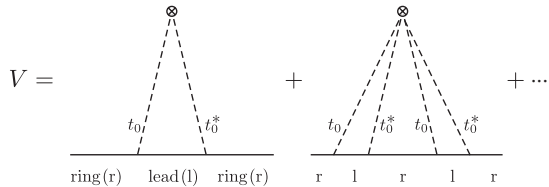}
\caption{Scattering off a contact to the lead with the initial and final states both in the ring. Summation in the full scattering amplitude $V$ is taken over all virtual transitions between the ring and the lead.}
\label{AppFig2}
\end{figure}

We calculate the amplitude of forward scattering off the contact, for the tunneling Hamiltonian, as illustrated in Fig.~\ref{AppFig2}. The total forward-scattering amplitude is given by a sum of the amplitudes of all virtual transitions between the ring and the lead \cite{aristov10}:
\be
V=-i\frac{2|t_0|^2/v}{1+2|t_0|^2/v^2}~.
\label{V}
\ee
When deriving Eq.~\eqref{V}, we assumed that the backscattering phase at the contact equals zero for an electron incident on the ring from the lead with momentum $k$, so that the wave function in the lead for $\gamma=0$ reads: $$\Psi(y)=\exp(iky)+\exp(-iky),\quad \Psi(y\to 0) \to 2$$ (here $y$ is the coordinate in the lead counted from the contact).

\begin{figure}[ht]
\includegraphics
[width=0.4\textwidth]{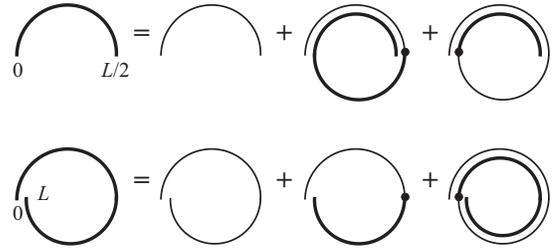}
\caption{Dyson equation for clockwise-propagating electrons. The thin and thick lines correspond to the bare and dressed Green functions, respectively.
The black dot denotes the forward scattering amplitude off a contact $V$.}
\label{AppFig3}
\end{figure}

The system of Dyson's equations for the Green functions at the contacts at $x=L/2$ and $x=L$ is depicted in Fig.~\ref{AppFig3} and reads
\BEA
G^+(L/2)&=&G_{0}^+(L/2)+G_{0}^+(L/2)VG^+(L)
\notag
\\
&+&G_{0}^+(L) VG^+(L/2)~,
\label{GL2}
\\
G^+(L)&=&G_{0}^+(L)+G_{0}^+(L/2)VG^+(L/2)
\notag\\
&+&G_{0}^+(L) VG^+(L)~.
\label{GL}
\EEA
(for the sake of brevity we omitted, only here, the argument $\epsilon$ and the index $R$ in the Green functions). The solution to Eqs.~\eqref{GL2} and \eqref{GL}, and to similar equations for counterclockwise-propagating electrons, is written as
\BEA
G_{R}^\pm(\epsilon,L/2)&=& -\frac{i}{v} \frac{e^{iq_\pm L/2}}{1-t_{\rm in}^2e^{iq_\pm L}}~, \label{GGpm2-dressed}
\\
G_{R}^\pm(\epsilon,L)&=& -\frac{i}{v} \frac{t_{\rm in}e^{iq_\pm L}}{1-t_{\rm in}^2e^{iq_\pm L}}~,
\label{GGpm-dressed}
\EEA
with $t_{\rm in}$ [given by Eq.~(\ref{tunnHam})] expressed in terms of $V$ as
\be
t_{\rm in}=1-iV/v~.
\label{t_in}
\ee

Now, note that Eqs.~\eqref{GGpm2-dressed} and \eqref{GGpm-dressed} for $\gamma\ll 1$ obey Eq.~\eqref{Green-eps-Gamma}. This is the property of the relation between the bare and tunneling-dressed Green functions that makes it useful in the study of tunneling in the presence of SCS. Specifically, as shown in Appendix \ref{properties}, the bare (no tunneling) Green function with SCS included is written in the vicinity of each of its poles in the form
\be
G_{R0}(\epsilon, x)\simeq\frac{a\exp (2\pi i M x/L)}{\epsilon-\varepsilon_{0}+i0}~,
\label{poles}
\ee
with a certain $x$ independent constant $a$ and a certain integer number $M$, both specific to the pole at $\epsilon=\varepsilon_0$. This is precisely the same form that the bare Green function has near its poles in the absence of SCS [Eq.~\eqref{pole}, with $a=1/L$ and $M=n$ near the pole at $\epsilon^\pm_n$]. As a result, the resummation of tunneling vertices by means of the Dyson equation results in Eq.~(\ref{Green-eps-Gamma}) for $\gamma\ll 1$ also in the presence of SCS.

\section{Green function of spinful electrons in a ring in the $(\epsilon,x)$ representation}
\label{properties}

In this Appendix, we derive the energy representation for the Green function of interacting spinful electrons in a ring. The function $\mathcal{G}_{SC}^+ $ [Eq.~\eqref{ga}], which describes SCS in a closed ring, can be rewritten as
\BEA
\mathcal{G}_{\rm SC}^+ (x,t) \!\!\!&=&\!\!\!\!\sum\limits_{nm}\!\int\limits_{-\infty}^{\infty}\!\int\limits_{-\infty}^{\infty}\frac{d\omega_1}{2\pi}\frac{d\omega_2}{2\pi}e^{-i\omega_1(t-x_n/u) -i\omega_2(t-x_m/v)}
\nonumber\\
\label{g2}
&\times&  \frac{\theta(t)}{2\pi^2 T \sqrt{uv}} \sum \limits_{\eta=\pm1} (- \eta) ~ a_{\omega_1}^\eta a_{\omega_2}^\eta~,
\EEA
where
\BEA
a_{\omega}^\eta &=& \int\limits_{-\infty}^{\infty}dt\frac{  \pi T~ e^{i\omega t}}{\sqrt{\sinh[\pi T(t-i\eta 0)]}}
\nonumber
\\
\nonumber
&=& \sqrt{\frac{16 i \pi }{\eta}} \,T\,\int\limits_{0}^{\infty}\int \limits_{0}^{\infty}dz dt \cos\left[\omega t- z^2 \eta \sinh(\pi Tt)\right]
\\
\nonumber
&=& \frac{\sqrt{2 i  \pi^3}}{\displaystyle \left |\Gamma\left(\frac34 +\frac{i\omega}{2\pi T} \right)\right|^2}~\frac{ \displaystyle \cosh \left(\frac{\omega}{2T}\right) + \eta~ \sinh \left(\frac{\omega}{2 T}\right) }{\sqrt{\eta }\displaystyle \cosh\left(\frac{\omega}{T}\right)}
\EEA
At $\omega=0$, we have
\be
a_0^\eta=\frac{\sqrt{2 i \pi^3}}{\sqrt{\eta}\,\Gamma^2(3/4)}~.
\ee
In the opposite limit of large $|\omega|$, using the asymptotic of the Gamma function
\be
\left |\Gamma\left( \frac34 +\frac{i\omega}{2\pi T} \right)\right|^2 \simeq \sqrt{\frac{2\pi |\omega|}{T}}
\exp\left(-\frac{|\omega|}{2T}\right), \quad \omega\to \pm \infty~,
\label{Gas}
\ee
$a_\omega^\eta$ behaves differently depending on the sign of $\omega\eta$:
\be
a_\omega^\eta\!=\!\sqrt{\frac{4 i \pi^2  T}{\eta |\omega|}}\times
\left\{
\begin{array}{lcl} \displaystyle
\!\!1,\quad  &\!\!\!\! \text{sgn}(\omega \eta)>0,\  |\omega|\to \infty~, \\ \displaystyle
\!\!\exp\!\left(\!-\frac{|\omega|}{ T} \right),\quad &\!\!\!\! \text{sgn}(\omega \eta)<0,\  |\omega|\to \infty~.
\end{array}
\right.
\nonumber
\ee
By applying the Poisson summation formula to Eq.~\eqref{g2}, we get
\begin{align}
&\mathcal{G}_{\rm SC}^+ (x,t) =-i\theta(t) \frac{\sqrt{\Delta \Delta_u}}{ T L}
\\
&\times \sum \limits_{n_1n_2}\!\!\ \lambda(n_1,n_2)e^{-i(n_1\Delta +n_2\Delta_u)t} e^{2\pi i x(n_1+n_2)/L}~,
\notag
\end{align}
where
\be
\lambda (n_1,n_2)=\frac{ 1}{4i\pi^3} \sum \limits_\eta~ \eta~ a_{n_1\Delta}^\eta ~ a_{n_2 \Delta_u}^\eta,
\label{lambda-nm0}
\ee
which gives Eqs.~\eqref{lambda-n1n2} and (\ref{lambda-nm}) of the main text.

The retarded Green functions in a closed ring are expressed, in the energy-coordinate representation, in terms of $\lambda (n_1,n_2)$ as follows:
\BEA
G_{R0}^{\pm}(\epsilon,x)&=&\frac{\sqrt{\Delta\Delta_u}}{TL}\sum\limits_{n_1n_2}\lambda (n_1,n_2)
\nonumber
\\
&\times& \frac{e^{2\pi i x(N_\pm +1+n_1+n_2)/L}}{\epsilon-\varepsilon({n_1,n_2})-\delta E^\pm_{\rm ZM} +i0}~,
\label{G0}
\EEA
where
\be
\varepsilon({n_1,n_2})=n_1\Delta +n_2\Delta_u \simeq \Delta (n+2\alpha m)
\ee
with $n=n_1+n_2$ and $m=n_2$ are the quantized energies of single-particle excitations in the presence of SCS [cf.\ Eq.~(\ref{fine-structure})]. It is worth noting that the Fourier transformation of Eq.~\eqref{G0} exactly factorizes into the ZM and SCS parts.

It is instructive to see how Eq.~\eqref{G0} transforms into the Green function of a noninteracting ring. To this end, we rewrite Eq.~\eqref{G0} for $\alpha=0$ as
\be
G_{R0}^{\pm}(\epsilon,x) =\frac{\Delta }{T L  }
\sum \limits_{nn_1}\frac{\lambda (n_1, n-n_1) ~e^{2\pi i x(N_\pm +1+n)/L} }{\epsilon-  n \Delta  - \delta E^\pm_{\rm ZM} +i0}
\label{G00}
\ee
By replacing the summation over $n_1$ with an integral and using Eq.~\eqref{prop}, we obtain
\be
G_{R0}^{\pm}(\epsilon,x) =\frac{1}{L} \sum \limits_{n}\frac{e^{2\pi i x(N_\pm +1+n)/L} }{\epsilon-  n \Delta  - \delta E^\pm_{\rm ZM} +i0}~.
\label{G00-non}
\ee
Substituting $\delta E^\pm_{\rm ZM}$ [Eq.~(\ref{dEZM})] for the case of zero interaction and shifting $n+N^\pm-N_0\to n$, Eq.~(\ref{G00-non}) gives the Green function of a noninteracting ring. The instructive point here is the importance of the ``sum rule" (\ref{prop}).

Finally, as was already mentioned at the end of Appendix \ref{App:Dyson}, we use the Dyson equation \eqref{Green-eps-Gamma} to find the Green function in the presence of SCS in the limit of $\gamma\ll 1$:
\begin{eqnarray}
{\bar G}_R^{\pm}(\epsilon,x)\!&=&\!\frac{\sqrt{\Delta \Delta_u}}{TL}
\label{G}
\\
&\times&\!\sum\limits_{n_1n_2}\frac{\lambda(n_1,n_2)\ e^{2\pi i x(N_\pm +1+n_1+n_2)/L}}{\epsilon-\varepsilon({n_1,n_2})-\delta E^\pm_{\rm ZM}+i\Gamma(n_1,n_2)/2}~,
\notag
\end{eqnarray}
where [cf.\ Eq.~(\ref{Gamma-nm})]
\be
\Gamma({n_1,n_2})= \Gamma_0\frac{\sqrt{\Delta\Delta_u}}{T}~\lambda({n_1,n_2}).
\label{Gn1n2}
\ee
In this way, we arrive at Eq.~(\ref{GR-pole}) of the main text.

\section{Dynamics of SCS beyond the $m$-approximation}
\label{App:C}

In this Appendix, we derive the functions $g(t)$, $K(t)$, and  $\tilde K(t)$ beyond the $m$-approximation introduced in Sec.~\ref{s3b}
[cf. Eqs.~(\ref{K18}), (\ref{gm}), (\ref{env}), and (\ref{envK0}) of the main text].
By Fourier transforming Eq.~\eqref{GR-pole} and using  Eqs.~\eqref{Green-finite}, \eqref{GZM},  and  \eqref{22}, we obtain
\be
g(t) =i \frac{\Delta\Delta_u}{2\pi T}\sum\limits_{nm} (-1)^{n+1} \lambda_{nm}e^{-i\varepsilon_{nm} t} e^{-\Gamma_{nm} t/2},
\label{g-exact}
\ee
where $\lambda_{nm}=\lambda(n-m,m)$.
The tunneling-induced decay for the sublevel $\varepsilon_{nm}$  is now characterized
by $\Gamma_{nm}$ [Eq.~(\ref{Gamma-nm})]
which is proportional to the structure factor $\lambda_{nm}$.
Next, substituting Eq.~(\ref{g-exact}) in Eq.~\eqref{12}, we find
 \begin{eqnarray}
 \nonumber
 K(t)&=&\frac{\Delta^2\Delta_u^2}{2\pi T^2} \sum\limits_{ nm, n'm' } (-1)^{n-n'}(-\p_\epsilon f)_{\epsilon=\varepsilon_{nm}}\lambda_{nm}\lambda_{n'm'}
 \\
 &\times&e^{-i(\varepsilon_{nm} -\varepsilon_{n'm'})t} e^{-(\Gamma_{nm} +\Gamma_{n'm'}) t/2}.
 \label{K-exact}
 \end{eqnarray}
The dependence of the structural factors $\lambda_{nm}$ on $n$ gives rise
to the double-horn structure of the peaks in $K(t)$ (see inset in Fig.~\ref{fig3})
instead of the delta functions of zero width as in Eqs.~(\ref{K18}) and (\ref{gm}).

In order to find the smooth envelope, $\tilde K(t)$, of the function $K(t)$,
we integrate Eq.~\eqref{K-exact} over the time interval $2\pi/\Delta$, which sets $n'=n.$
As a result, we find that $\tilde K(t)$ in a closed ring (i.e., for $\Gamma_{nm}=0$) is expressed
as a single sum
\BEA
\nonumber
\tilde K(t)&=&\left \langle \left|  \frac{\Delta }{ T} \sum \limits_{m} \lambda(n-m,m)  e^{-2 i \alpha \Delta t m}  \right|^2  \right \rangle_n
 \\
\label{tildeK}
&=& \frac{\Delta}{T} \sum\limits_k b_k e^{-2i \alpha \Delta k t}.
\EEA
Here, $\langle \ldots \rangle_n$ stands for $ -(\Delta/T)\sum\limits_n  \left[\ldots (\p_\epsilon f )_{\epsilon=n\Delta}\right]$
and the coefficients $b_k$ are related to $\lambda_{nm}$ as
\be
b_k = \frac{\Delta^2}{T} \sum \limits_{n,m} \lambda_{nm} \lambda_{n,m-k}  \left(-\p_\epsilon f\right)_{\epsilon=n\Delta}.
\label{ak}
\ee

Equation \eqref{ak} can be further simplified for $T/\Delta\gg 1$ by replacing the summation over $(n,m)$ with integrals.
In this way, we get
\be
b_k\simeq \mathcal{B}\left(\frac{k\Delta}{T}\right),
\label{bB}
\ee
where
\be
\mathcal{B}(z)\!=\!\int\!dE dy\Lambda(E-y,y)\Lambda(E-y+z, y-z)(-\p_E f  )
\label{AKL}
\ee
with $E=\epsilon/T$ and $\Lambda(x_1,x_2)$ given by Eq.~(\ref{lambda-nm}).
Substituting Eq.~\eqref{lambda-nm} in Eq.~\eqref{AKL},
we see that $E$ enters Eq.~(\ref{AKL}) only in the combination $E-y$.
Shifting the integration variable $E\to E-y$ and using  Eq.~\eqref{prop}, we obtain
\be
\mathcal{B}(z)=\frac{1}{4\cosh^2(z/2)}.
\label{BB}
\ee
Replacing the sum over $k$ in Eq.~\eqref{tildeK} with an integral and using Eqs.~\eqref{bB} and \eqref{BB},
we obtain the function $I(z)$ [Eq.~\eqref{37a}].

Finally, let us explain which property of the structural factor $\lambda_{nm}$
accounts for the double-horn structure of the peaks in $g(t)$ and $K(t)$.
Consider, for definiteness, a peak in $K(t)$ at $t=t_*=2\pi\left(n_*+1/2\right)/\Delta$
belonging to the bunch with $k=0$ (a generalization to $k\neq 0$ is straightforward).
Assume that the peak is close to the center of the bunch, so that its width $(\Delta t)_{n_*n_*}$
is much smaller than $1/T$, as discussed in Sec.~\ref{s3a}.
For such a narrow peak, Eq.~\eqref{g-theta} can be written explicitly as
\be
g(t) \simeq -i\,\frac{\theta[\delta t (2\alpha t_*-\delta t )]}{\pi \sqrt{\delta t (2\alpha t_* -\delta t)}}~,
\label{g-horns}
\ee
where $\delta t=t-t_*$.
Note that $T$ drops out from Eq.~(\ref{g-horns}).
Thus, the double-horn structure of the narrow peaks close to the centers of the bunches is
the same as in the limit $T\to 0$.

Let us now demonstrate that Eq.~(\ref{g-horns}) is determined by the asymptotic behavior of $\Lambda (x_1,x_2)$ [Eq.~(\ref{lambda-nm})] for large values of its arguments:
\be
\Lambda (x_1,x_2)\simeq\frac{\displaystyle e^{|x_1+x_2|/2-|x_1|/2-|x_2|/2}}{\pi\sqrt{|x_1x_2|}}~,\quad |x_{1,2}|\gg 1~.
\label{Lam-Asympt}
\ee
From Eq.~(\ref{Lam-Asympt}) we find that the function $\lambda_{nm}$ decays exponentially as a function of $m$ outside the interval $-|n|<m<|n|$,
on a characteristic scale of $|m-n|\sim T/\Delta$.
Neglecting these exponential tails for $|n|\gg T/\Delta$, we find for $\lambda_{nm}$ in this limit:
\be
\lambda_{nm}\simeq\frac{T}{\pi\Delta}\frac{\theta[m(n-m)]}{\sqrt{m(n-m)}}~.
\label{lambda-nm-large}
\ee
Within this approximation, $m$ is limited to the interval $(0,n)$ for $n>0$ and to the interval $(n,0)$ for $n<0$.
Substituting Eq.~\eqref{lambda-nm-large} in Eq.~(\ref{g-exact}), which represents $g(t)$ in terms of $\lambda_{nm}$,
putting $\Gamma_{nm}=0$ and only keeping  $\alpha\ll 1$ in $\varepsilon_{nm}$, we have
\be
g(t)\simeq -i\theta(t)\frac{\Delta^2}{2\pi T}\sum\limits_{nm}(-1)^n \lambda_{nm}e^{-i\Delta (n+2\alpha m) t }~.
\ee
In the vicinity of the $n_*$-th peak, we obtain
\be
g(t)\simeq -i\frac{\Delta}{2\pi^2 }\sum\limits_{nm}\frac{\theta[m(n-m)]}{\sqrt{m(n-m)}} e^{i\Delta(  n \delta t  - 2\alpha m t_*) }~.
\ee
Replacing the sums over $n$ and $m$ with integrals, we reproduce Eq.~\eqref{g-horns}.
The double-horn structure (\ref{g-horns}) of the peaks in $g(t)$ can thus be seen to be directly related to the
double-horn dependence of $\lambda_{nm}$ on $m$ in Eq.~\eqref{lambda-nm-large}.

\section{Conductance beyond the $m$-approximation}
\label{beyond-m}

In this Appendix, we calculate the conductance through the ring within the isotropic model
without neglecting the dependence of $\lambda_{nm}$ on $n$
(thus going beyond the $m$-approximation discussed in Sec.~\ref{s3b}).
Such a calculation is more conveniently performed in the energy-coordinate representation, by substituting Eq.~\eqref{GR-pole} in Eq.~\eqref{15}.
For $\alpha\ll 1$, we neglect the difference between $u$ and $v$ everywhere except for the energies $\varepsilon_{nm}$ [Eq.~(\ref{fine-structure})]
in the resonant denominators of the Green functions.

Consider first the classical term in the conductance, which involves the product of two Green functions with the same chirality
and is given by the sum over $(n,m)$ and $(n',m')$. For $\alpha \Delta \gg \Gamma$, the broadening of the sublevels
$\Gamma_{nm}$ [Eq.~(\ref{Gamma-nm})] is smaller
than the distance between them. Neglecting all the terms in the sum except those with  $n=n'$ and $m=m'$, we get
\BEA
{\cal T}_c&=& \left(\frac{ \Delta \Gamma_0}{2 T}\right)^2\int  d\epsilon \left(-\p_\epsilon f\right)\\ \nonumber
&\times &\sum \limits_{nm\pm}\left \langle \frac{\lambda_{nm}^2}{(\epsilon -\varepsilon_{nm} -\delta E^\pm_{\rm ZM})^2+ (\Gamma_{nm}/2)^2} \right \rangle_{\rm  \!\! ZM} .
\label{Tc}
\EEA
The thermal factor $\left(-\p_\epsilon f\right)$ is a smooth function on the scale of a typical width $\Gamma_0\Delta/T$ of the Lorentz peaks in Eq.~\eqref{Tc},
so that it can be taken at the positions of the sublevels in the integral over $\epsilon$.
Further, for $\Delta/T\ll 1$
we can write for each sublevel
$$(\p_\epsilon f)|_{\epsilon=\varepsilon_{nm} -\delta E^\pm_{\rm ZM}}\simeq (\p_\epsilon f )|_{\epsilon =\Delta n }.$$
Using Eq.~\eqref{Gn1n2}, we obtain
\be
{\cal T}_c=\frac{\pi \Delta \Gamma_0}{ T}  \sum\limits_{nm} \lambda_{nm} (-\p_\epsilon f)|_{\epsilon=n\Delta}.
\label{sum}
\ee
Replacing the sums over $n$ and $m$ with integrals and using Eq.~\eqref{prop}, we arrive at Eq.~\eqref{Drude} of the main text.

The quantum contribution to the transmission coefficient is calculated in a similar manner.
Substituting Eq.~\eqref{GR-pole} in Eq.~\eqref{15}, we now take the product of the Green functions of opposite chiralities:
\BEA
&&\hspace*{-0.3cm}{\cal T}_q=\pi \left(\frac{ \Delta \Gamma_0}{T}\right)^2 {\rm  Re} \sum \limits _{nn'mm' } i \lambda_{nm} \lambda_{n'm'}
\label{58a}
\\
\nonumber
&&\hspace*{-0.3cm}\times
\left \langle \frac{  (-1)^{N_+-N_- +n-n'} (-\p_\epsilon f)_{\epsilon=n\Delta}}{
\delta E^+_{\rm ZM}-\delta E^-_{\rm ZM} +\varepsilon_{nm}-\varepsilon_{n'm'}+i\Gamma_\varphi}
\right \rangle_{\rm \!\! ZM}~.
\EEA
Here we introduced a phenomenological dephasing rate $\Gamma_\varphi$ in the denominator and
neglected the sublevel broadening $\Gamma_{nm}$, assuming that $\Gamma_\varphi\gg\Gamma_{nm}$.
Using
\be
\delta E^+_{\rm ZM}-\delta E^-_{\rm ZM} = \Delta (N_+-N_-  -1) - 2\Delta  \delta \phi
\label{C*}
\ee
for the difference of the ZM energies and Eq.~(\ref{fine-structure}) for $\varepsilon_{nm}$, we write
\BEA
&&\hspace*{-0.3cm}{\cal T}_q=\pi \left(\frac{ \Delta \Gamma_0}{T}\right)^2 {\rm  Re} \sum \limits _{nn'mm' } i \lambda_{nm} \lambda_{n'm'}
\label{58}
\\
\nonumber
&&\hspace*{-0.3cm}\times
\left \langle \frac{  (-1)^{\bar N} (-\p_\epsilon f)_{\epsilon=n\Delta}}{\Delta (\bar N-1) -2\Delta  \delta \phi +2\alpha \Delta (m-m')+i \Gamma_\varphi }
\right \rangle_{\rm \!\! ZM}
\EEA
with $\bar N=N_+-N_- +n-n'$.

Since $\alpha \ll 1$ and $\Delta\gg \Gamma_\varphi$,
we can retain in the sum in Eq.~(\ref{58}) only the terms corresponding to $\bar N=1$ (i.e., $n'=n+N_+-N_- -1$).
[In the coordinate-time representation, this corresponds to the averaging of $K(t)$
over the period $2\pi/\Delta$ which leads to the replacement of $K(t)$ with a smooth envelope $\tilde K$.]
Therefore, $(-1)^{\bar N}=-1,$ and, consequently, the sign of ${\cal  T}_q$
is opposite to that of ${\cal  T}_c$, which implies destructive interference.

Further, since $\lambda_{nm}$ changes on the scale $\delta n  \gtrsim T/\Delta$,  while the  fluctuations of   $N_+-N_-$ around zero
are of the order of $\sqrt{T/\Delta}\ll \delta n $, one can approximate $\lambda_{n+N_+-N_- -1,m}\simeq \lambda_{nm}$.
The ZM numbers then drop out, and the interference part of the transmission coefficient is given by a single sum over narrow Lorentz peaks:
\be
{\cal T}_q= - \frac{ \pi \Gamma_0^2 }{ T} \sum\limits_k  b_k ~\frac{\Gamma_\varphi}{ 4\Delta^2 (\delta \phi -\alpha k)^2+\Gamma_\varphi^2}
\label{T-int-1}
\ee
with the coefficients $b_k$ defined in Eq.~(\ref{ak}).
The sum of Eq.~(\ref{T-int-1}) and Eq.~(\ref{Drude}) gives the final form of the conductance
in the isotropic model [Eq.~(\ref{finalG}) of the main text].


\begin{thebibliography}{100}
\bibitem{fractionalization}  Equation (\ref{ga}) is written within the square-root approximation. A more general expression, which corresponds  to Eq.~\eqref{18} in free space
and includes  fractionalization, reads, for $T\gg \Delta$:
\BEA
{\cal G}_{\rm SC}^+
&\simeq &\frac{i\theta(t)}{ \pi\sqrt{u v}D^{\alpha_b/2}}\,{\rm Im}\sum\limits_{nml}A^{\alpha_b/4}_u(x_l,-t+i0)
\nonumber\\ \nonumber
&\times& A^{1/2+\alpha_b/4}_u(x_n,t-i0)A^{1/2}_v(x_m,t-i0)~.
\EEA
From this equation, one can see that fractionalization does not manifest itself in our problem.  Indeed, we have three independent summations over  $n,m$, and $l$.  It follows that, when considering a maximum of, say, $A_u(x_n,t-i0)$ at the contact  with $x_n=L(n+1/2)$,  there exists $l$  such  that $A^{\alpha_b/4}_u(x_l,-t+i0)\simeq 1$, which implies ``unfractionalization'' of  the electron charge at the contact. This, in turn, means that fractionalization does not change  the single-particle spectrum. In particular, Fig.~\ref{F-new} remains  intact.
\bibitem{footnote-contacts} In the interacting ring, the tunneling amplitudes are renormalized by Luttinger-liquid effects. Importantly, within the square-root approximation for the Green functions, there is no renormalization of the pointlike tunneling. As a consequence, the renormalization of tunneling contacts is essentially the same as in the spinless case, see Ref.~\cite{aristov10}. In what follows, we include this $T$ dependent renormalization into the definition of the tunneling amplitudes.
\bibitem{asymmetric} One can generalize this result to the setup with different tunneling rates of two contacts, $\Gamma_1$ and $\Gamma_2$. In this case, one should replace $\Gamma_0\to\sqrt{\Gamma_1\Gamma_2}$ in the prefactor of the conductance in Eq.~(\ref{finalG}) and $\Gamma_0\to (\Gamma_1+\Gamma_2)/2$ everywhere in the square brackets, including the expression for $\Gamma_\varphi$ in Eq.~\eqref{Gamma-phi}.
\bibitem{footnote-ZM} We relegate the discussion of other conceivable mechanisms of ``intrinsic'' dephasing in the symmetric cases (i.e., in the absence of the otherwise dominant ZM dephasing) to future work, assuming in this paper that the phenomenological external dephasing is stronger than the (if existing, suppressed compared to generic ZM dephasing) intrinsic one.
\end{thebibliography}
\end{document}